# A theoretic analysis of magnetoactive GES-based turbulent solar plasma instability

Souvik Das,[1] Ahmed Atteya[2] and Pralay Kumar Karmakar [1]★

[1]*Department of Physics, Tezpur University, Napaam, Assam 784028, India*
[2]*Department of Physics, Faculty of Science, Alexandria University, Alexandria 21511, Egypt*

## ABSTRACT

A recently reported gravito-electrostatic sheath (GES) model is procedurally applied to study the turbumagnetoactive helioseismic oscillation features in the entire bi-fluidic solar plasma system. The bounded solar interior plasma (SIP, internally self-gravitating), and the unbounded solar wind plasma (SWP, externally point-gravitating) are coupled through the interfacial diffused solar surface boundary (SSB) due to an exact gravito-electrostatic interplay. A numerical platform on the developed theoretic formalism reveals the evolution of both dispersive and non-dispersive features of the modified GES mode fluctuations in new parametric windows. Different colourspectral profiles exhibit important features of the GES-based SIP–SWP perturbations elaborately. It is illustratively shown that the thermostatistical GES stability depends mainly on the radial distance, magnetic field, equilibrium plasma density, and plasma temperature. We see that their dispersive features are more pertinently pronounced in the self-gravitational domains (SIP) than the electrostatic counterparts (SWP). Besides, different characteristic parameters with accelerating (or decelerating) and stabilizing (or destabilizing) effects influencing the entire solar plasma stability are illustratively portrayed. We speculate that, in the SIP, the long-wave (gravitational-like) helioseismic fluctuations become highly dispersive showing more propagatory nature than the shorter ones (acoustic-like). The short waves show more propagatory propensity than the longer ones in the SSB and SWP regime. The reliability of our proposed investigation is bolstered along with the tentative applicability and future scope in light of the current solar observational scenarios, such as SOHO, STEREO, SDO, PSP, and SolO.

**Key words:** instabilities – Sun: helioseismology – Sun: interior – Sun: oscillations – Solar wind.

## 1 INTRODUCTION

The Sun, like all other living stars in galaxies, is basically a bounded structure in nature because of its inherent self-gravitational confinement mechanisms against the outward acoustic pressure (Christensen-Dalsgaard 2002; Gurnett et al. 2002; Aschwanden 2014; Kasper et al. 2021). Such processes, leading thereby to the mysterious development of the solar surface boundary (SSB), are yet to be well understood from a collective plasma-wall interaction perspective. It eventually implicates that the plasma boundary wall-interaction mechanism plays an important role, inhabitable in principle, for consideration and subsequent actualization in studying the solar plasma stability dynamics in a real sense. The intrinsic magnetic field present in the Sun and its circumambient atmosphere acts as a vital energy source and affects all the inherent properties defining the self-gravitationally confined solar plasma dynamics (Priest 2014; Goutam & Karmakar 2015; Brun & Browning 2017). This inhomogeneous magnetic field is internally produced by the collective convective circulation dynamo action of the charged particles constituting the entire solar plasma system (Narita 2012; Priest 2014; Brun & Browning 2017). It is noteworthy that the entire solar plasma system is actually non-static and non-uniform in nature. There exist a plethora of collective waves, oscillations, and fluctuations in the entire volume of the solar plasma system (Hale et al. 2022). These waves influence various inherent solar plasma parameters. Understanding those solar waves and oscillations could provide a solid diagnostic tool in probing the structural and dynamical properties of solar interior and atmosphere (Christensen-Dalsgaard 2002; Hansen et al. 2004; Aschwanden 2014; Bhattacharya & Hanasoge 2023). In this context, probing the solar internal structure in light of the collective solar plasma modes and oscillations has developed a growing interest among the astro-plasma communities.

The study of diversified helioseismic modes has been one of the captivating topics enabling us to understand the interior structural properties of the Sun and its atmosphere. It is already a well-known fact that, helioseismology provides a fine diagnostic interpretation of the diverse characteristics of the collective oscillation and wave modes, excited in the entire solar plasma system (Christensen-Dalsgaard 2002; Hansen et al. 2004; Kosovichev 2006a, 1999; Aschwanden 2014; Priest 2014; Ambastha 2020; Gizon et al. 2021). It has been successful to extract new insights to study the physical properties of the solar interior structure mainly. These helioseismic waves and oscillations are produced on the SSB due to stochastically excited sound waves, generated and trapped inside the solar convective zone, resulting from the pressure fluctuations and turbulent convective motions (Kosovichev 2006a). These waves penetrate the solar interior, return to the SSB, and also to the solar exterior. It hereby registers relevant information on the diverse morphodynamical features of the internal structure of the Sun and its complex interlayer coupling dynamics (Duvall et al. 1993; Kosovichev 2006a; Ambastha 2020; Mani et al. 2022). This is because this category of collective waves carries forward an important glimpse about the basic structural features of the media they pass through collectively. As a consequence, a thorough decoding analysis of the same in such hot solar environs throws light to various solar structural aspects, long-lying inaccessible directly, in reality, to any kind of direct experimental measurements as far as seen widely.

Three distinct types of collective modes produced by helioseismic activities in the solar plasma system are thermal pressure-driven acoustic mode (p mode), internal gravity-driven gravity mode (g mode), and buoyancy-driven surface gravity mode (f mode) (Christensen-Dalsgaard 2002; Hansen et al. 2004; Kosovichev 2006a; Ambastha 2020; Gizon et al. 2021). Thus, it is evident that the restoring factors behind these distinct modes are thermal pressure, gravitational pressure, and buoyant pressure, respectively. It is expedient to mention here that the p-mode amplitude increases from the SSB outwards; whereas, that of the g mode only inwards against the SSB towards the core region. In contrast, the f-mode amplitude is actively dominant only in the SSB region, and not elsewhere. The p mode gets excited stochastically by the convective turbulence mechanism effect beneath the photosphere layer and damped by radiative losses (Harvey et al. 1996; Christensen-Dalsgaard 2002; Stix 2002; Gizon et al. 2021). As a consequence, large-scale plasma flow energy (gravitational, $K \to 0$; $K$ : Jeans-normalized angular wavenumber, also illustratively explained later) gets cascaded collectively to short-scale energy spectrum (electrostatic, $K \to \infty$). It is herewith pertinent to add that the helioseismic p-mode characterization plays an important role to probe the morphodynamics of the solar interior structure extensively. It is against the remaining modes (g-, f-) affected technically from observational constraints, such as weak modal amplitude and lack of detection resolution (Christensen-Dalsgaard 2002; García & Ballot 2019; Gizon et al. 2021).

It is worth mentioning here that solar astrophysicists have already reported several observational studies on the solar 5-min oscillations. This, in fact, explicitly reveal diversified resonant modes of acoustic waves (p-mode spectrum) travelling (velocity $\sim 1.5 \times 10^{-1}$ m s$^{-1}$) from the SSB to the solar interior towards the centre of the entire solar plasma mass distribution (Leighton et al. 1962; Ulrich 1970; Deubner 1975; Rhodes, Ulrich & Simon 1977; Demarque & Guenther 1999; Stix 2002; Aschwanden 2014; Gizon et al. 2021). In other words, it manifests the local acoustic spectral components of the global p-mode phenomenon pressure-excited in the solar plasma system.

In addition to the above, a fourth category of helioseismic modes, which is indeed a hybrid gp mode composed of the g mode (interior) and p mode (exterior) coupled via the SSB, could also exist in the sun-like stars away from the main sequence in the Hertzsprung–Russell (HR) diagram (Stix 2002; Priest 2014; Ambastha 2020). Significant astronomic signatures in support of the existence of such hybrid modes have been well confirmed by several observational missions, such as the VLT-ULTRACAM, Convection Rotation and Planetary Transits space mission (CoRoT), Kepler, and so forth (García & Ballot 2019; Aerts 2021; Buldgen et al. 2022). However, remarkable explorations in this hybrid gp direction from the weak-amplitude asteroseismic perspective are yet to be carried out precisely.

Diversified collective modes naturalistically excited by solar convection enable astrophysicists to understand the interior dynamics of the Sun and its atmosphere. A complete identification and characterization of the solar plasmic modes in the framework of standard solar models (SSMs) has still been lying as an open challenge. At this emerging backdrop, a highly relevant exploration in this plasma-based helio-physical direction has been the study of the diversified linear and non-linear modal dynamics of collective solar plasma fluctuations. The local mode excitation and fluctuation mechanism naturally are due to the intrinsic parametric perturbations triggered by diverse local mechano-thermic and non-local self-gravity disturbances (Stix 2002; Hansen et al. 2004; Aschwanden 2014; Priest 2014; Ambastha 2020).

In the laboratory-produced plasmas, a thin non-neutral space charge layer, called the plasma sheath, is formed near the plasma-confining wall (Chen 1984; Bittencourt 2004). In order for this sheath structure to form, a local criterion on the threshold ionic-flow at the sheath entrance, termed as the Bohm sheath criterion ($M > 1$), is to be fulfilled (Oksuz & Hershkowitz 2005; Davood Sadatian & Gharjeh ghiyaei 2021). In analogy with the laboratory-bounded plasma systems, a gravito-electrostatic sheath (GES) structure has been reported to be formed at the gravitational potential boundary, called the SSB, because of widely separated gravitothermal coupling effects of the constitutive electrons and ions on the solar plasma scales (Dwivedi et al. 2007; Karmakar et al. 2016). According to this GES model, the entire solar plasma system is divided into two scales coupled through the diffused SSB (Dwivedi et al. 2007) given as (i) solar interior plasma (SIP, bounded) and (ii) solar wind plasma (SWP, unbounded).

It is noteworthy that, similar to the laboratory-bounded plasmas, there exists an equivalent form of the local Bohm sheath criterion on the threshold ionic escape velocity ($M_\infty \geq \sqrt{2}$) needed for the formation of this quasi-neutral isothermal GES on the solar scales (Karmakar & Dwivedi 2011). However, the Bohm sheath criterion for a far-from thermal equilibrium (i.e. non-thermal) plasma system has been obtained as: $M > [(\kappa - d_{e,\phi} + 1)/\kappa]^{1/2}$; where, $d_{e,\phi}$ ($\ll 1$) is the number of degrees of freedom of the constitutive electrons, dependent on the perturbation potential. Here, $\kappa$ is the non-thermality spectral (Kappa) index (Livadiotis 2015; Davood Sadatian & Gharjeh ghiyaei 2021). In non-thermal plasma conditions, $0 < \kappa < 1$. Hence, we find the ion escape velocity across the gravitational potential barrier as $M_\infty \geq \sqrt{2}$. The strength of the self-gravitational potential wall is such that the massive ions (colder) cannot overcome it at the SSB regime. On the other hand, the lighter electrons (hotter) can escape the SSB potential barrier. As a consequence, there occurs a thermal leakage of the electron flow against the ions at the SSB, thereby developing an electrostatic polarization-induced astrophysical GES. The solar self-gravitational plasmas with radial variation of self-gravity exhibit its maximum potential strength at the SSB zone (Karmakar et al. 2016). It is noteworthy that the self-gravitationally confined solar plasmas are radially inhomogeneous and non-uniform in nature due to the presence of diversified inherent zeroth-order microphysical gradient forces against the laboratory ones (Priest 2014; Karmakar et al. 2016).

It is extensively seen that the bounded structure formation processes in astrophysical fluids need a local outward halting support against the inward self-gravitational collapsing force. The fluid turbulence plays an important role in exciting diverse wave-fluctuation patterns responsible for the transport of mass momentum energy among constitutive fluid elements (Narita 2012). It thereby leads to the development of hydrostatically bounded equilibrium structures (Stix 2002; Hansen et al. 2004; Priest 2014). It is pertinent to mention here that, as the gravitational Poisson equation is time-independent in analytic construct; therefore, the Sun and Sun-like stellar structures remain self-gravitating even after being so formed in a bounded configuration. As a consequence, the inclusion of turbulence and its effects on the self-gravity action is important in studying the GES-based instability dynamics to explore the collective modes excitable in the Sun and its ambient complex atmosphere. It has been reported in an earlier investigation (Goutam & Karmakar 2016) in this context that, due to the turbumagnetic pressure effects, the sheath width increases by 5.17% relative to the ideal GES structure.

A semi-analytical study of non-thermal GES model fluctuation dynamics modified by the q-non-extensive non-thermal electrons and non-linear logabarotropic pressure effects has been reported earlier (Gohain & Karmakar 2018). It has shown therein that the lowest order GES stability depends exclusively on the electron–ion temperature ratio ($T_e/T_i$) and the radial position coordinate ($\xi$). The damping behaviour of the fluctuations is more relevant in the acoustic-like domain than the gravitational one (Gohain & Karmakar 2018).

We have recently employed a polytropic macroscopic state formalism founded in a bi-fluidic theory describing both the constitutive electrons and ions to explore the GES-based instability featuring the magnetoactive solar plasmas in a helioseismic modal prospective (Das & Karmakar 2022). It has systematically applied a normal spherical mode analysis without any traditional quasi-classic approximation. However, it has been restricted to simplified dispersive illustrations only in a restricted parametric pattern. The present contribution puts forward a continued GES-centric theoretic study on the similar helioseismic modal fluctuations and instabilities. It is sourced purely in self-gravitating plasma-wall interaction processes in more realistic multiparametric spectral regimes elaborately in the context of the GES-based solar plasma stability dynamics previously remaining fully unaddressed and unexplored.

A considerable number of solar observations have provided significant data on collective helioseismic oscillations propagating in the solar plasma system. Some main data sets are provided by several ground-based networks, such as the Taiwan Oscillation Network (TON), the Global Oscillation Network Group (GONG), and Birmingham Solar Oscillation Network (BiSON; Kosovichev 1999; Christensen-Dalsgaard 2002; Stix 2002; Casanellas et al. 2012; Ambastha 2020; Gizon et al. 2021). Various space telescopes, like the Michelson Doppler Imager (MDI) instrument in the SOlar and Heliospheric Orbiter (SOHO) spacecraft launched jointly by the European Space Agency (ESA) and the National Aeronautics and Space Administration (NASA); and the Helioseismic and Magnetic Imager (HMI) onboard the NASA-operated Solar Dynamics Observatory (SDO) have identified diversified modes of collective oscillations in the solar surface (Christensen-Dalsgaard 2002; Stix 2002; Kosovichev 2006a; Casanellas et al. 2012; Ambastha 2020; Cohen et al. 2021; Gizon et al. 2021). The role of plasma instabilities in shaping the dynamics of solar plasma through magnetoactive expansion effects has recently been reported (Kellogg 2022). This study is based on various observational data source, including the NASA's Parker Solar Probe (PSP), Solar Terrestrial Relations Observatory (STEREO), Coordinated Data Analysis Web (CDAWeb), and so forth.

Motivated by the above-mentioned scenarios, we develop a theoretic model formalism to investigate helioseismic excitation mode and fluctuation dynamics in the entire magnetoactive GES-based solar plasma system. Hence, we implement a spherical normal mode analysis ($\xi \nrightarrow \infty$) to see the active helioseismic stability features associated with both the bounded (SIP) and unbounded (SWP) scales. In our model, we consider the inclusion of polytropic effect, turbumagnetic action, non-planar fluctuations with an exclusion of any kind of conventional quasi-classic approximation judiciously, so forth. Our focused key novelty is to formulate GES-based spherical normal mode helioseismic fluctuation dynamic study in a thermal bi-fluidic model fabric. Unlike the previously reported planar ($\xi \to \infty$) investigations describing a simplified plane-wave approximation study (Gohain & Karmakar 2018), our speculation reveals that the modified non-planar GES-based helioseismic fluctuations do not only exhibit a traditional radial ($\xi$) dependence. The fluctuations are also affected simultaneously by the average solar plasma density ($n_0$), magnetic field ($B_0$), and solar core-to-electron temperature ratio ($T_0/T_e$) on both the bounded SIP and unbounded SWP scales sensibly.

Apart from the introduction part, the structural layout of this manuscript is organized in a standard pattern as follows. Section 2 deals with the physical model and mathematical formalism of the solar plasma system. The SIP and SWP analyses are described in Sections 2.1 and 2.2, respectively. The results and discussion are depicted in Section 3. The SIP and SWP scale outcomes are illustrated in Sections 3.1 and 3.2, respectively. The atypical dimensional p-mode characteristic features investigated here are illustratively interpreted and validated in Section 3.3. Finally, the main conclusions drawn from our semi-analytical investigation alongside tentative future scope in light of the diversified existing astronomical observations in a pure helioseismic perspective are summarily presented in Section 4.

## 2 PHYSICAL MODEL AND FORMALISM

We consider a simple bi-fluidic quasi-neutral turbumagnetoactive solar plasma system in the spherically symmetric GES-based model (radial, 1D, reduced degrees of freedom) consisting of mainly ionized form of hydrogen (92%) and helium (8%) (Stix 2002; Aschwanden 2014; Priest 2014). We ignore many other heavy ionic and neutral species, such as $\alpha$-particles, C, N, O, Fe, etc. It is because of their poor relative abundance (about 0.01%) in the solar atmosphere (Stix 2002; Hansen et al. 2004; Priest 2014; Ambastha 2020). This yields a simplified analysis to anticipate a dynamical depiction of the entire solar plasma fluctuation mode associated with it. We portray a schematic diagram (Fig. 1) of the Sun and its circumambient plasma atmosphere according to (a) SSM and (b) GES for the sake of comparative conceptualization of the readers. It is depicted that the solar surface in the former (SSM) lies in the photosphere layer from where most of the Sun's energy is emitted as light; while, the surface in latter (GES) is defined by the maximization of the solar self-gravity wall strength. The localized solar self-gravitational potential barrier with radial variation of its own acts as a non-rigid enclosure to confine this quasi-neutral plasma with its maximum strength at the derived SSB regime ($\xi \approx 3.5$) (Dwivedi et al. 2007). The entire solar plasma system, embedded in an inhomogeneous magnetic field, dynamically couples the subsonic bounded SIP and supersonic unbounded SWP scales via the interfacial SSB formed under the non-local action of the long-range gravito-electrostatic force field (Dwivedi et al. 2007; Karmakar et al. 2016). The constitutive thermal electronic and ionic fluids are coupled via the gravito-electrostatic Poisson formalism on the solar plasma scales of space and time. The fluid turbulence effect arises here due to the overlapping of multiple micro-kinematical scales of the constitutive species (Narita 2012; Goutam & Karmakar 2016). In our model, such fluid turbulence effects are modelled by the Larson logabarotropic equation of state. The effects of any kind of non-ideal, gyro-fluidic, and tidal actions are ignored hereby.

The global quasi-neutral nature ($n_e \approx n_i = n$) of the entire solar plasma system is justifiably based on the realistic ground that the asymptotic value of the Debye-to-Jeans length ratio is almost zero ($\lambda_D/\lambda_J \approx 10^{-20} \sim 0$) (Dwivedi et al. 2007; Goutam & Karmakar 2015). It is noteworthy that after forming via the *Jeansean* dynamic molecular cloud collapse mechanism, the Sun (SIP) continues to remain as a self-gravitating system in nature because of the time-independent nature of the self-gravitational Poisson equation in this non-relativistic approach. This fact strengthens the reliability of the gravitational Poisson equation applied in our model and justifiably enables us to employ the *Jeansean* spatiotemporal scales for the adopted standard astrophysical normalization (re-scaling in a dimensionless form) scheme.

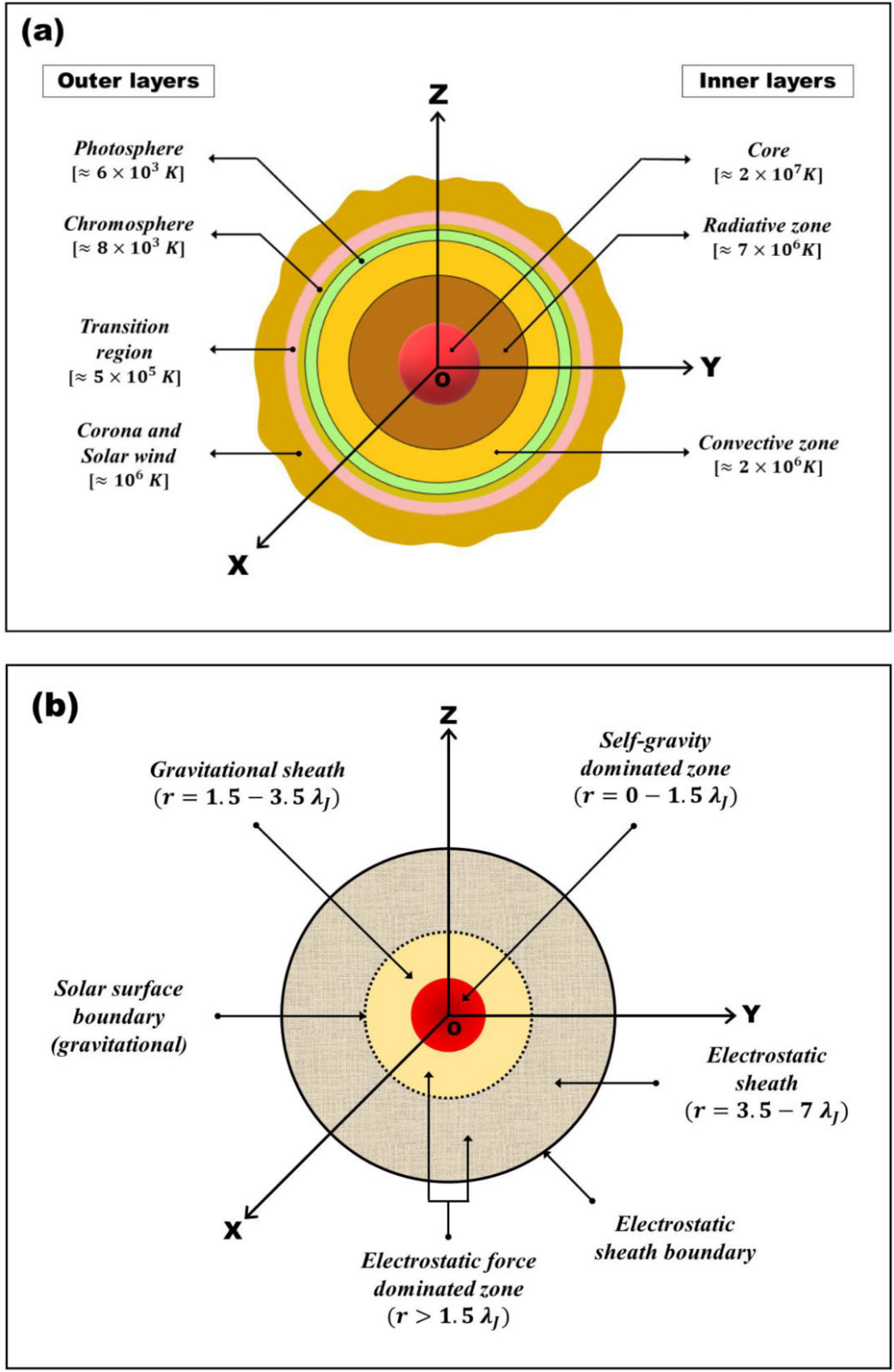

**Figure 1.** Schematic diagram of the Sun and its circumvent atmosphere according to the (a) SSM and (b) GES model. Different concentric constitutive layers of the models are depicted separately in a similar footing. The radius of the Sun (photospheric radius) as per the SSM is comparable to the solar surface boundary (SSB, solar radius) as per the proposed GES model theory.

According to the virial theorem, the outward thermo-electrostatic coupling force balances the inward self-gravitational force and prevents the Sun to undergo further self-gravitational collapse (Stix 2002; Priest 2014; Brun & Browning 2017). It is pertinent to add here that the origin of the thermal pressure force lies in the contraction effects and that of the electrostatic counterpart is in the collective plasma dynamics leading to the Coulombic expansion. Thus, the Sun maintains a hydrostatic equilibrium state, thereby retaining its main-sequence stage. Hence, the self-gravitational subsonic SIP (non-Newtonian, non-point source) is bounded by the self-gravitational Poisson equation. However, for the unbounded supersonic SWP scale (Newtonian, point source), the SIP itself acts as an external gravity source, thereby making the solar self-gravitational Poisson equation redundant in this SWP context.

It is noticeable that most of the astrophysical bounded structures and circumambient atmospheres are naturally driven to re-organization by large-scale non-local gravito-electrostatic coupling caused by long-range interparticle force exhibiting non-extensivity (non-Maxwellian) property. Such non-extensive systems are governed by Tsallis thermostatistical framework, which is, indeed, a generalization of the Boltzmann–Gibbs (BG) statistics (Jiulin 2006, 2007; Livadiotis & McComas 2009, 2013; Livadiotis 2015; Aman-Ur-Rehman & Lee 2018). A quasi-statistical equilibrium condition, using the Tsallis $q$-entropic index, is given as $q = 1 + (\vec{\nabla} T / \vec{\nabla} G)$; where $T$ and $G$ denote the temperature (in energy units) and the gravitational potential energy, respectively (Jiulin 2006, 2007; Gohain & Karmakar 2018). Thus, in the special case, when $\vec{\nabla} T = \vec{0}$ i.e. at thermalized state with constant thermodynamic potential, we get $q = 1$ which reduces the thermostatistical distribution function to the well-known Maxwell–Boltzmann (MB) statistics (Jiulin 2006; Livadiotis & McComas 2009; Aman-Ur-Rehman & Lee 2018). In collisionless space plasmas, in addition to the Tsallis $q$-non-extensive distribution, there exists another pertinent non-thermal thermostatistical distribution called the Kappa distribution ($\kappa$) to govern the non-Maxwellian suprathermal behaviour of the constitutive particles (Livadiotis & McComas 2009, 2013; Livadiotis 2015). The thermal Maxwellian distribution actually is a special case of the Kappa distribution in the asymptotic limiting case of $\kappa \to \infty$ (Rubab & Murtaza 2006; Livadiotis & McComas 2009, 2013; Livadiotis 2015). The entropic index $q$ (used in non-extensive statistics) and the non-thermality spectral (Kappa) index $\kappa$ (used in astrospace inhomogeneous plasmas) hold a suitable relationship under the inter transformational equation (Livadiotis & McComas 2009, 2013; Livadiotis 2015; Aman-Ur-Rehman & Lee 2018) as $\kappa = 1/(q - 1)$ or, $q = 1 + (1/\kappa)$. In a non-Maxwellian Kappa modified plasma model, the Debye length differs from the traditional Debye length which is valid only in the Maxwellian plasmas. The standard relation between the Kappa modified Debye length ($\lambda_{D\kappa}$) and the Maxwellian Debye length ($\lambda_D$) is cast (Rubab & Murtaza 2006) as $\lambda_{D\kappa} = \lambda_D[(\kappa - 1.5)/(\kappa - 0.5)]^{1/2}$. Applying the well-known $\kappa - q$ transformation equation as mentioned earlier, we obtain the equivalent Debye length equation in $q$-form as $\lambda_{Dq} = \lambda_D[(2.5 - 1.5q)/(2 - 0.5q)]^{1/2}$.

It is well-known that astrophysical plasmas, including the entire solar plasma system, exhibit polytropic behaviour, governed by a generalized polytropic equation of state. The polytropism is incorporated through a specific mathematical relationship between the plasma fluid density and temperature in a hydrostatic equilibrium configuration (Hansen et al. 2004; Priest 2014). Hence, we employ a simple polytropic equation of state for the effective pressures, $p = K_p \rho^\gamma = K_p \rho^{(1 + n_p^{-1})}$; where, $K_p$ denotes the polytropic constant, $\gamma = 5/3$ denotes the polytropic exponent, and $n_p = (\gamma - 1)^{-1}$ denotes the corresponding polytropic index (Hansen et al. 2004; Narita 2012; Priest 2014; Vidotto 2021). Moreover, the active turbulence pressure, present in the solar plasma system, is given by a logabarotropic equation of state, $p_{turb} = p_0 \log(\rho/\rho_c)$; where $p_0$ is the mean (equilibrium) pressure (thermal), $\rho$ denotes the material volumetric density, and $\rho_c$ represents its heliospheric core value (Vazquez-Semadeni et al. 1998; Narita 2012). The effects of polytropic, turbulent (logabarotropic), and magnetic pressures conjointly act on the complex solar plasma system towards the re-organization of a new equilibrium set-up to be well understood (Goutam & Karmakar 2016).

To carry out our scale-free calculation scheme, various standard significant notations and symbols associated with the adopted standard normalization scheme (Christensen-Dalsgaard 2002; Goutam & Karmakar 2016; Gohain & Karmakar 2018) are given in Appendix A.

Applying all the customary notations and symbols from Appendix A pertinent to the helio-plasma dynamics (Dwivedi et al. 2007; Gohain & Karmakar, 2015, 2018; Goutam & Karmakar 2016), we describe our adopted standard astrophysical normalization scheme with all the usual *Jeansean* notations and significances in Appendix B.

The entire solar plasma system is governed by the continuity equation, momentum equation, electromagnetic induction equation, self-gravitational Poisson equation (for SIP only), and electrostatic Poisson equation (Dwivedi et al. 2007). Thus, the basic set of the governing equations in unnormalized form are cast with all the customary notations, respectively, as

$$\partial_t n_{e(i)} + \vec{\nabla} \cdot \left( n_{e(i)} \vec{v}_{e(i)} \right) = 0, \tag{1}$$

$$m_{e(i)} n_{e(i)} \left[ \partial_t \vec{v}_{e(i)} + \left( \vec{v}_{e(i)} \cdot \vec{\nabla} \right) \vec{v}_{e(i)} \right] = q_{e(i)} n_{e(i)} \vec{E} - \vec{\nabla} p_{e(i)T} + m_{e(i)} n_{e(i)} \vec{g}, \tag{2}$$

$$\partial_t \vec{B} = \vec{\nabla} \times \left( \vec{v}_{e(i)} \times \vec{B} \right), \tag{3}$$

$$\nabla^2 \psi = 4\pi G \rho, \tag{4}$$

$$\nabla^2 \phi = -\rho_q/\epsilon_0 = e \left( n_e - n_i \right)/\epsilon_0 \tag{5}$$

The generalized unnormalized forms of the above equations (equations 1–5) in the radial (1D) form with all the usual notations for the spherically symmetric GES-based solar plasma description can be recast, respectively, as

$$\partial_t n_{e(i)} + (r^{-2})\partial_r \left( r^2 n_{e(i)} v_{e(i)} \right) = 0, \tag{6}$$

$$m_{e(i)} n_{e(i)} \partial_t v_{e(i)} = -q_{e(i)} n_{e(i)} \partial_r \phi - \partial_r p_{e(i)T} - m_{e(i)} n_{e(i)} \partial_r \psi, \tag{7}$$

$$B_{az} \partial_r v_{e(i)} + r^{-1} B_{az} v_{e(i)} = -\partial_t B_{az}, \tag{8}$$

$$\partial_r^2 \psi + 2r^{-1}\partial_r \psi = 4\pi G m_i n_i, \tag{9}$$

$$\partial_r^2 \phi + 2r^{-1}\partial_r \phi = e\epsilon_0^{-1} \left( n_e - n_i \right) \tag{10}$$

As already mentioned, all the notations used in the above set of equations are generic. $\partial_{t(r)}$ denotes here the partial differentiation operator with respect to time (distance). It could be further mentioned that only the azimuthal component of the solar magnetic field is considered in our proposed calculation scheme.

Applying the relevant astrophysical normalization scheme (Appendix B), we now proceed systematically to present our GES-based helioseismic analyses for both the subsonic SIP (bounded) and supersonic SWP (unbounded) scales separately.

## 2.1 The SIP scale analysis

The self-gravitationally bounded SIP configuration is considered as a weakly magnetized collisional plasma system governed by the well-known set of the Jeans-normalized helio-structuring equations (Goutam & Karmakar 2015). The same set of the SIP equations (equations 1–5) in the normalized form are reproduced and cast, respectively, as

$$\partial_\tau N_{e(i)} + M_{e(i)}\partial_\xi N_{e(i)} + N_{e(i)}\partial_\xi M_{e(i)} + (2/\xi)\,N_{e(i)}\,M_{e(i)} = 0, \quad (11)$$

$$\begin{aligned}
\partial_\tau M_{e(i)} = {} & s\,\partial_\xi \Phi - N_{e(i)}^{-1}\partial_\xi N_{e(i)}\left[1 + \epsilon_{T_0}N_{e(i)}^{-1}\right] \\
& - N_{e(i)}^{-1}\alpha\,B_{az}^*\partial_\xi B_{az}^* - \partial_\xi \Psi,
\end{aligned} \quad (12)$$

$$B_{az}^*\left(\partial_\xi M_{e(i)} + \xi^{-1}M_{e(i)}\right) + \partial_\tau B_{az}^* = 0, \quad (13)$$

$$\partial_\xi^2 \Psi + 2\xi^{-1}\partial_\xi \Psi = N_i, \quad (14)$$

$$(\lambda_D/\lambda_J)^2\left[\partial_\xi^2 \Phi + 2\xi^{-1}\,\partial_\xi \Phi\right] = N_e - N_i = 0 \quad (15)$$

In equation (12), the term $s$ denotes the electrostatic polarity phase factor with values $s = +1$ (for electrons) and $s = -1$ (for ions). $\epsilon_{T_0} = T_0/T_e$ represents the solar core-to-electron temperature ratio. The $(2/\xi)$-term originates due to the consideration of spherical geometry ($\xi \to \infty$); which would, otherwise, be absent in the case of plane parallel geometry ($\xi \nrightarrow \infty$). This term causes a decrease in the ion density flux radially outwards. This happens because the radial distance ($\xi$) increases with the intervening surface area of the spherical surfaces. This effect is termed as the 'geometrical curvature effect' in the solar plasma flow dynamics (Karmakar et al. 2016). We have already mentioned the fact that the solar plasma system exhibits global quasi-neutrality behaviour ($N_e \approx N_i = N$) in nature, which, indeed, is clearly evident from the electrostatic Poisson equation for the potential distribution (equation 15).

The main source of the collective helioseismic disturbances excited in the entire solar plasma system is attributable to the diversified myriad acoustic waves, originated and trapped in the turbulent convection zone, located just beneath the solar surface in the purview of the standard solar models (Stix 2002; Priest 2014; Ambastha 2020). Probing the solar interior morphostructure diagnostically through the solar asteroseismology (helioseismology) hitherto is concentrated on the study of the diversified normal mode frequencies of the collective solar oscillations (Duvall et al. 1993; Christensen-Dalsgaard 2002). These eigenmodes associated with the Sun and its surrounding atmosphere behave as standing wave patterns, confined within a resonant cavity, bounded by the solar surface. The wave propagation depth of the so-formed solar cavity resonator is purely subject to the wavenumbers and the frequencies of the spectral eigenmode oscillations in the system (Christensen-Dalsgaard 2002; Mani et al. 2022).

The goal here is to study the GES-based helioseismic instability dynamics of the entire spherically symmetric solar plasma volume.

No polar and azimuthal counterparts are considered herein for the sake of simplicity. It applies a local (linear) spherical (radial) normal mode perturbation (small amplitude) analysis (Karmakar et al. 2016; Gohain & Karmakar 2018; Das & Karmakar 2022) against the well-defined hydrostatic homogeneous equilibrium configuration in the customary symbolism as

$$F(\xi, \tau) = F_0 + F_1(\xi, \tau), \quad (16)$$

$$F = \left[N_e\, N_i\, M_e\, M_i\, \Phi\, \Psi\, B_{az}^*\right]^T, \quad (17)$$

$$F_0 = [1\,1\,0\,0\,0\,0\,1]^T, \quad (18)$$

$$F_1 = \left[N_{e1}\, N_{i1}\, M_{e1}\, M_{i1}\, \Phi_1\, \Psi_1\, B_{az1}^*\right]^T \quad (19)$$

Here, $F_0$ denotes the set of non-perturbed (equilibrium) values of the relevant solar plasma parameters. $F_1(\xi, \tau) \sim (1/\xi)\,e^{-i(\Omega\tau - K\xi)}$ represents the corresponding perturbations arising due to the helioseismic effects. Here, $F_1(\xi, \tau)$ is also called the Eulerian perturbation at a given spatial point (Christensen-Dalsgaard 2002). This considered non-planar perturbations evolve as restricted spherical waves with the Jeans-normalized angular frequency $\Omega$ and the Jeans-normalized angular wavenumber $K$. Equation (16) enables us to transform the direct coordination space ($\xi, \tau$) into the reciprocal Fourier space ($K, \Omega$). It leads to commensurable transformation of linear differential operators from the direct radial coordinates to the inverted wave coordinates as $\partial/\partial\tau \to -i\Omega$, $\partial/\partial\xi \to (iK - 1/\xi)$, and $\partial^2/\partial\xi^2 \to (2/\xi^2 - K^2 - 2iK/\xi)$. An orderly application of equations (16)–(19) in equations (11)–(15) in the ($\xi, \tau$)-space results in their corresponding algebraic versions on the relevant physical perturbation in the ($K, \Omega$)-space cast, respectively, as

$$M_{e1(i1)}\left(iK + \xi^{-1}\right)\; - i\Omega N_{e1(i1)} = 0, \quad (20)$$

$$\begin{aligned}
i\Omega M_{e1(i1)} = {} & \left(iK - \xi^{-1}\right) \\
& \times \left[s\Phi_1 + N_{e1(i1)} + \epsilon_{T_0}N_{e1(i1)} + \alpha B_{az1}^* + \Psi_1\right],
\end{aligned} \quad (21)$$

$$\Omega B_{az1}^* = M_{e1(i1)}K, \quad (22)$$

$$\Psi_1 = -N_{i1}/K^2, \quad (23)$$

$$N_{e1} = N_{i1} \quad (24)$$

Equations (20)–(24) describe the oscillation due to helioseismic perturbation on the self-gravitationally bounded SIP scale. Performing the elimination procedure, equations (20)–(24) finally decouple into a linear generalized quadratic dispersion relation in the Fourier space ($K, \Omega$) for the SIP fluctuation dynamics cast with all the generic notations as

$$\Omega^2 = (K^2 + \xi^{-2})\left[1 + \epsilon_{T_0} + \alpha K(K^2 + \xi^{-2})^{-1}(K + i\xi^{-1}) - K^{-2}\right] \quad (25)$$

Using composite frequency $\Omega = (\Omega_r + i\Omega_i)$ in equation (25) and following quadratic equation solving method, we obtain the real (normal and regular) and imaginary (perturbed and irregular) frequency parts characterizing the GES-based SIP instability modes, respectively, as

$$\Omega_r(\xi, K) = 2^{-1/2}\left[\chi + \left(\chi^2 + \alpha^2 K^2\xi^{-2}\right)^{1/2}\right]^{1/2}, \quad (26)$$

$$\Omega_i(\xi, K) = -2^{-1/2}\left[-\chi + \left(\chi^2 + \alpha^2 K^2\xi^{-2}\right)^{1/2}\right]^{1/2}; \quad (27)$$

where

$$\chi\left(\xi, K\right) = \left(K^2 + \xi^{-2}\right)\left[1 + \epsilon_{T_0} + \alpha K^2\left(K^2 + \xi^{-2}\right)^{-1} - K^{-2}\right] . \quad (28)$$

Equations (26)–(27) explicitly show that our proposed non-planar turbumagnetoactive bi-fluidic GES-based SIP stability dynamics significantly depends on the Jeans normalized radial distance ($\xi$), solar core-to-electron temperature ratio ($\epsilon_{T_0}$), and the magneto-thermal pressure coupling constant or magneto-thermal pressure coupling parameter ($\alpha$). The novelty of our study is multidimensionally justifiable due to the inclusion of all these significant terms and factors in the basic model set-up. It especially deals with various $\alpha$-scenarios in our spherical wave analysis in diversified thermal configurations. This proposed investigation clearly differs from the previously reported solar stability analyses founded on the plane geometry consideration clubbed with the help of the conventional quasi-classic asymptotic approximation of the parametric fluctuations (Gohain & Karmakar 2018).

An exact analytical shape matching with the pre-reported planar analysis based on the plane-wave ($\xi \to \infty$) approximation (Gohain & Karmakar 2018) concretizes the reliability of our derived non-planar ($\xi \nrightarrow \infty$) GES-based SIP dispersion relation (equation 25).

### 2.2 The SWP scale analysis

It is well known that the hot solar corona generates suprathermal high energetic particles. The solar wind is the continuous supersonic outflow of completely ionized gas i.e. plasmas, streaming away from the solar corona (Parker 1958; Stix 2002; Priest 2014). The SWP is an unbounded collisionless magnetized turbulent plasma system that excites plasma microinstabilities because of its transition from collision-dominated (SIP-region) to collisionless behaviour and exhibits a large number of collective plasma oscillation phenomena (Dwivedi et al. 2007; Priest 2014). In this unbounded SWP scale, the self-gravity is switched off and get transformed to external gravity as the SIP plays as an external Newtonian point source. This yields in the redundance of the gravitational Poisson equation in SWP scale (Dwivedi et al. 2007; Karmakar & Dwivedi 2011; Goutam & Karmakar 2016). Therefore, the relevant normalized electron(ion) momentum equations governing the SWP dynamics are orderly given with all the generic notations as

$$\partial_\tau M_{e(i)} = s\partial_\xi\Phi - N_{e(i)}^{-1}\partial_\xi N_{e(i)}\left[1 + \epsilon_{T_0}N_{e(i)}^{-1}\right] \\ -\alpha B_{az}^{*-1}N_{e(i)}^{-1}\partial_\xi B_{az}^{*} - a_0\xi^{-2} . \quad (29)$$

Here, the term $a_0 = GM_\odot/c_s^2\lambda_J = 95$ is used as a normalization coefficient which provides an estimated measurement of SWP temperature. All the other remaining equations (continuity, magnetic induction, and electrostatic Poisson) describing the SWP constitutive species have the same mathematical form as derived earlier in the case of SIP scale.

Applying the same Fourier transformation method and equations (16)–(19), we obtain the same set of algebraic expression (as equations 20, 22, and 24) in the $(K, \Omega)$-space describing the linear perturbation features of the SWP dynamics. The SWP momentum equations (equation 29) with the relevant linear perturbations is reproduced and cast as

$$i\Omega M_{e1(i1)} = \left(iK - \xi^{-1}\right)\left[s\Phi_1 + N_{e1(i1)} + \epsilon_{T_0}N_{e1(i1)} + \alpha B_{az1}^{*}\right] \\ +2a_0\xi^{-2}N_{e1(i1)} . \quad (30)$$

Employing the same elimination and decoupling method over the perturbed equations governing the unbounded SWP stability dy-namics, we derive a linear generalized quadratic dispersion relation describing the oscillation features of the SWP scale cast as

$$\Omega^2 = \left(K^2 + \xi^{-2}\right)\left[1 + \epsilon_{T_0} + \alpha K\left(K^2 + \xi^{-2}\right)^{-1}\left(K + i\xi^{-1}\right)\right] \\ -2a_0\xi^{-2}\left(iK + \xi^{-1}\right) . \quad (31)$$

Applying $\Omega = \left(\Omega_r + i\Omega_i\right)$ in equation (31) and performing the same root-finding method, we obtain the pertinent set of roots $(\Omega_r, \Omega_i)$ presented, respectively, as

$$\Omega_r\left(\xi, K\right) = 2^{-1/2}\left[\zeta + \left\{\zeta^2 + \left(\alpha K\xi^{-1} - 2a_0K\xi^{-2}\right)^2\right\}^{1/2}\right]^{1/2} , \quad (32)$$

$$\Omega_i\left(\xi, K\right) = -2^{-1/2}\left[-\zeta + \left\{\zeta^2 + \left(\alpha K\xi^{-1} - 2a_0K\xi^{-2}\right)^2\right\}^{1/2}\right]^{1/2} ; \quad (33)$$

where

$$\zeta = \left[\left(K^2 + \xi^{-2}\right)\left\{1 + \epsilon_{T_0} + \alpha K^2\left(K^2 + \xi^{-2}\right)^{-1}\right\} - 2a_0\xi^{-3}\right] . \quad (34)$$

It is observed from equations (32) and (33) that the turbumagnetoactive GES-based SWP stability features depends mainly on the Jeans normalized radial distance ($\xi$), solar core-to-electron temperature ratio ($\epsilon_{T_0} = T_0/T_e$), and the magneto-thermal pressure coupling constant ($\alpha$) in an analogous fashion as we found previously for the case of SIP regime as well.

In a similar way, the validation of our non-planar ($\xi \nrightarrow \infty$) GES-based SWP dispersion relation is justifiably concretized with an exact analytic shape-matching method with the previously reported planar analysis grounded on the well-known plane-wave ($\xi \to \infty$) approximation (Gohain & Karmakar 2018).

## 3 RESULTS AND DISCUSSION

The helioseismic oscillation features of the turbumagnetoactive GES-based bi-fluidic SIP and SWP regimes are analytically developed. Two distinct sets of basic governing equations in normalized form are formulated for both the bounded subsonic SIP and unbounded supersonic SWP scales. Employing non-planar perturbation analysis, a pair of generalized quadratic dispersion relations is obtained for both the SIP (equation 25) and SWP (equation 31) scales. In our analysis, instead of the quasi-classic short-wavelength plane-wave approximation ($K\xi \gg 1$, or $\xi \gg \lambda$) as reported earlier (Gohain & Karmakar 2018), we carry out GES-centric spherical wave perturbation analysis. As already mentioned, the quasi-linear bi-scale relationship is structurally and analytically established through the coupling via the long-range non-local GES force field action (Dwivedi et al. 2007; Karmakar & Dwivedi 2011; Goutam & Karmakar 2016; Karmakar et al. 2016; Gohain & Karmakar 2018). In other words, this autocoupling arises as the solar interior self-gravity (non-Newtonian) converts into an external point gravity source (Newtonian) via the diffused interfacial SSB. Both the dispersion relations are theoretically and numerically analysed in the framework of expedient set of input values relevant to the solar plasma system. Incorporating other relevant suitable solar parameters, we obtain graphical colourspectral profiles (Figs 2–28) of collective helioseismic waves on both the SIP and SWP spatiotemporal regimes. We now proceed to the analytic discussions of the stability features on both the bounded SIP and unbounded SWP scales elaborately.

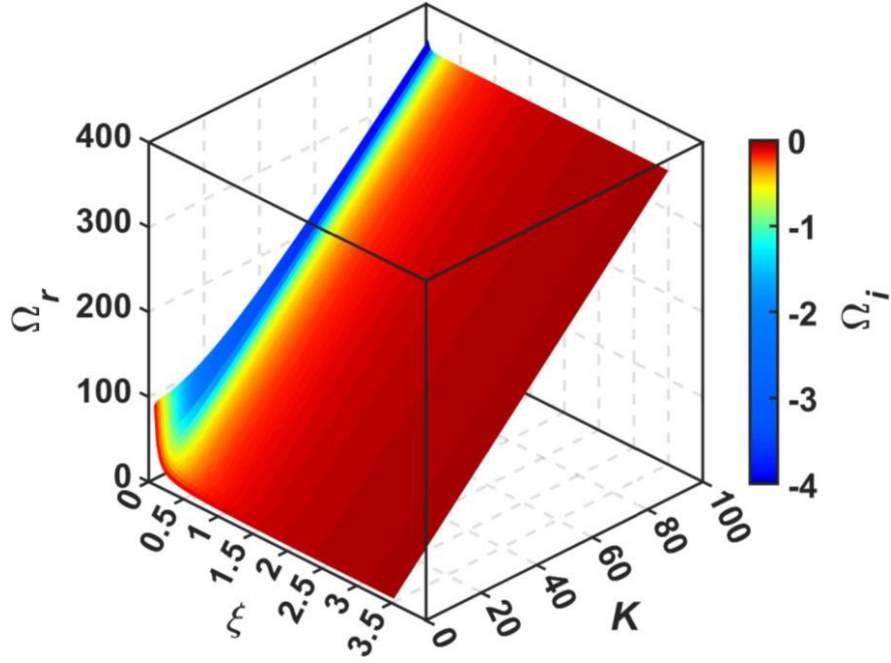

**Figure 2.** Colourspectral profile of the Jeans-normalized real frequency ($\Omega_r$) and imaginary frequency ($\Omega_i$) jointly with variation in the Jeans-normalized radial distance ($\xi$) and the Jeans-normalized angular wavenumber ($K$) associated with the SIP fluctuation dynamics.

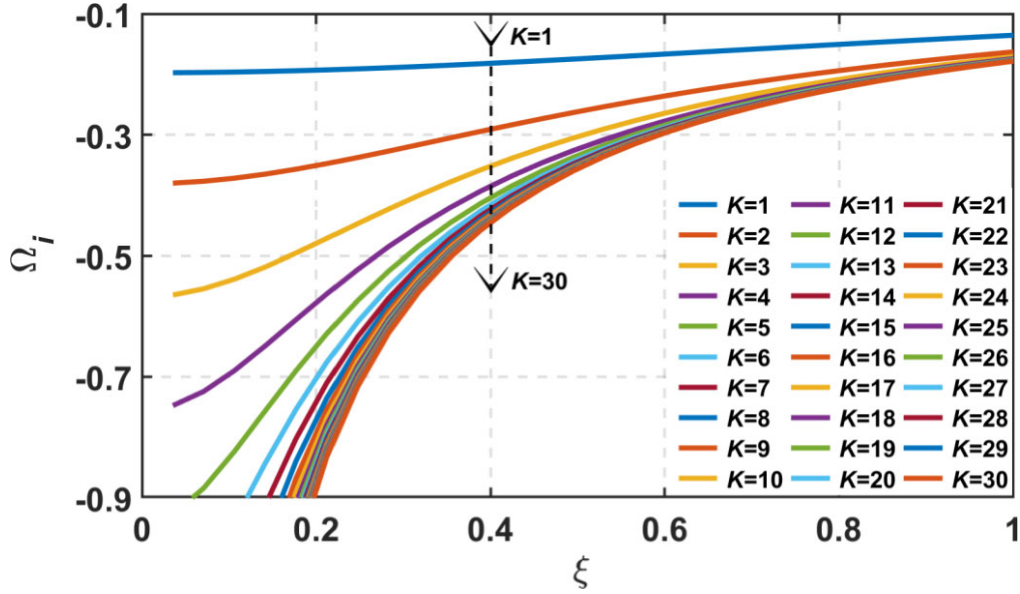

**Figure 3.** Profile of the Jeans-normalized imaginary frequency part ($\Omega_i$) with variation in the Jeans-normalized radial distance ($\xi$) for different indicated values of Jeans-normalized angular wavenumber ($K$) associated with the SIP fluctuation dynamics.

### 3.1 SIP stability outcomes

The Jeans-normalized real frequency part ($\Omega_r$) and imaginary frequency part ($\Omega_i$) expressed, respectively, in equations (26) and (27) are numerically analysed in detail. Using the suitable input values relevant to the SIP scale, we obtain graphical patterns of the SIP stability features as portrayed in Figs 2–13. In Fig. 2, we depict the colourspectral profile of the Jeans-normalized real ($\Omega_r$) and imaginary ($\Omega_i$) frequencies with the variation in the Jeans-normalized radial distance ($\xi$) from the centre of the entire SIP mass distribution outwards and associated Jeans-normalized angular wavenumber ($K$). The magneto-thermal pressure coupling constant is kept fixed here at $\alpha = 1.25$. We also fix the input values of the solar core and SIP electron temperatures (in energy units), assuring an exact hydrostatic force balancing in order to form the SSB, respectively, as $T_0 = 10^3$ eV and $T_e = 10^2$ eV (Dwivedi et al. 2007; Goutam & Karmakar 2016); thereby, adopting the solar core-to-SIP electron temperature ratio as $\epsilon_{T_0} = T_0/T_e = 10^3/10^2 = 10$. It is clearly observed that near the vicinity of the solar core ($\xi \sim 0$) for $K \leq 40$ (i.e. for longer waves), the SIP fluctuations show dispersive nature; while, beyond that region, it becomes purely acoustic-like non-dispersive in nature. It is well confirmed by the real linear dispersion relationship $\Omega_r = f(K)$. It is also found that, beyond the core region ($\xi > 0.1$), $\Omega_r$ remains constant with $\xi$. It signifies that, in the SIP regime, the helioseismic fluctuations show stable oscillatory propagation. In contrast, the $\Omega_i$-magnitude decreases (i.e. damping rate decreases) in the $\xi$-space as we go outwards from the centre of the entire SIP mass distribution indicating instability propensity of the oscillations (Fig. 2).

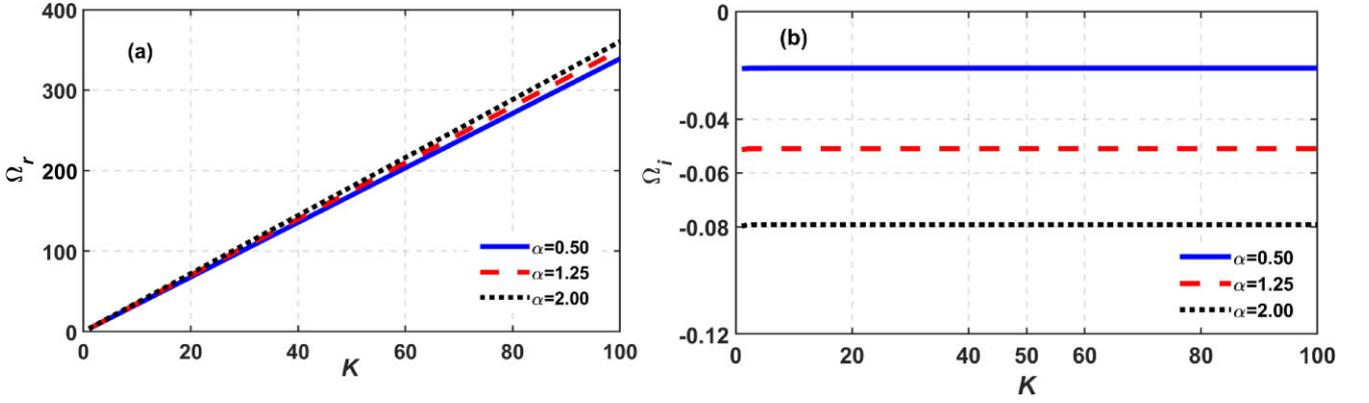

**Figure 4.** Profile of the Jeans-normalized (a) real frequency ($\Omega_r$) and (b) imaginary frequency ($\Omega_i$) with variation in the Jeans-normalized angular wavenumber ($K$) for different indicated values of the magneto-thermal pressure coupling constant ($\alpha$) associated with the SIP fluctuation dynamics.

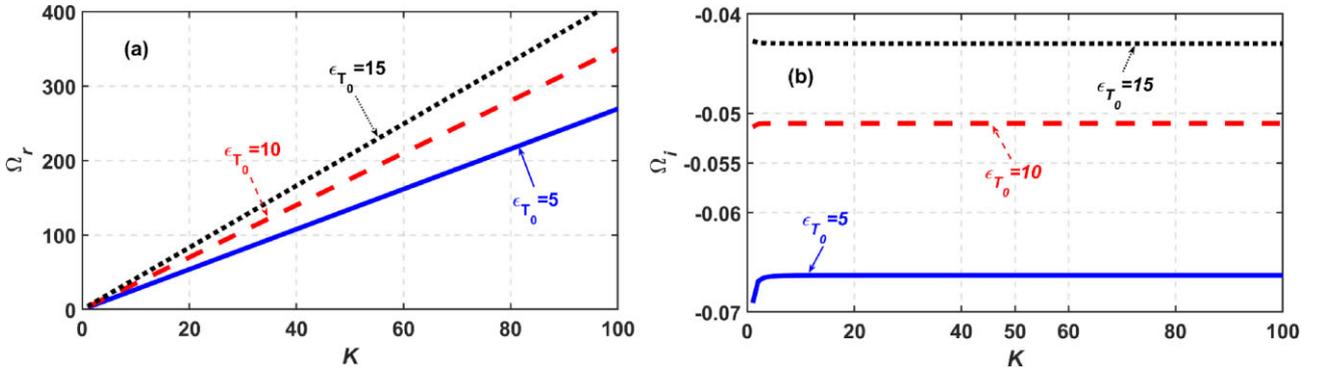

**Figure 5.** Profile of the Jeans-normalized (a) real frequency ($\Omega_r$) and (b) imaginary frequency ($\Omega_i$) with variation in the Jeans-normalized angular wavenumber ($K$) for different indicated values of the solar core-to-electron temperature ratio ($\epsilon_{T_0}$) associated with the SIP fluctuation dynamics.

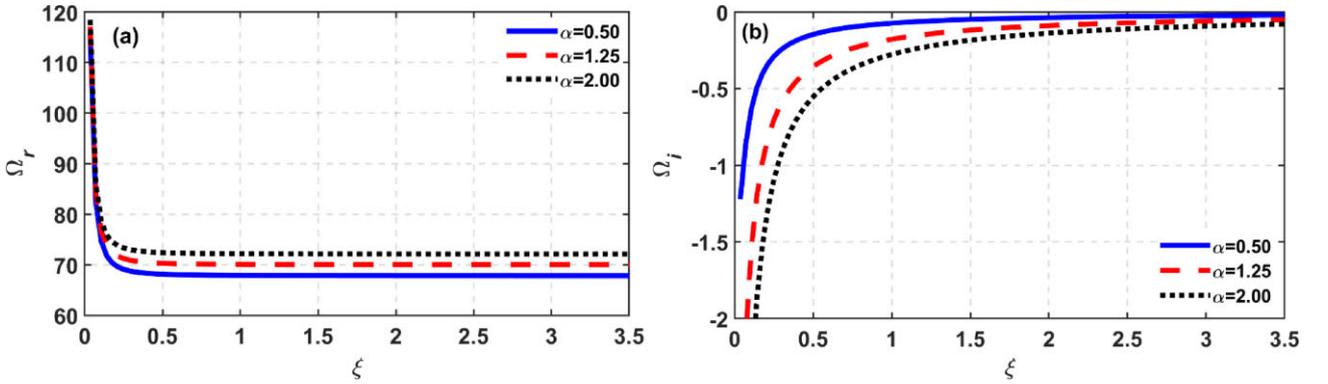

**Figure 6.** Profile of the Jeans-normalized (a) real frequency ($\Omega_r$) and (b) imaginary frequency ($\Omega_i$) with variation in the Jeans normalized radial distance ($\xi$) for different indicated values of the magneto-thermal pressure coupling constant ($\alpha$) associated with the SIP fluctuation dynamics.

In Fig. 3, we portray the profile of $\Omega_i$ with variation in $\xi$ for different indicated values of $K$ associated with the SIP fluctuation dynamics. It illustrates the damping features of the SIP oscillations. It is interestingly seen that, near the solar core ($\xi < 0.4$), the damping rate of the shorter wavelength fluctuations (acoustic-like) is much more prominent than that of the longer wavelength (gravitational-like) ones. These collective waves, oscillations, and fluctuations are found to be more damped near the solar core than in the outward radial zones exhibiting the SIP fluctuations away from the core region (Fig. 3).

Fig. 4 displays the profiles of (a) $\Omega_r$ and (b) $\Omega_i$ with variation in $K$ for different values of $\alpha$ (e.g. $\alpha = 0.50$, 1.25, 2.00) for the SIP fluctuation dynamics on the Jeans spatial scale at the SSB region ($\xi \sim 3.5$). It is seen that, at the SSB, the SIP perturbations show a non-dispersive acoustic-like nature holding a linear relationship

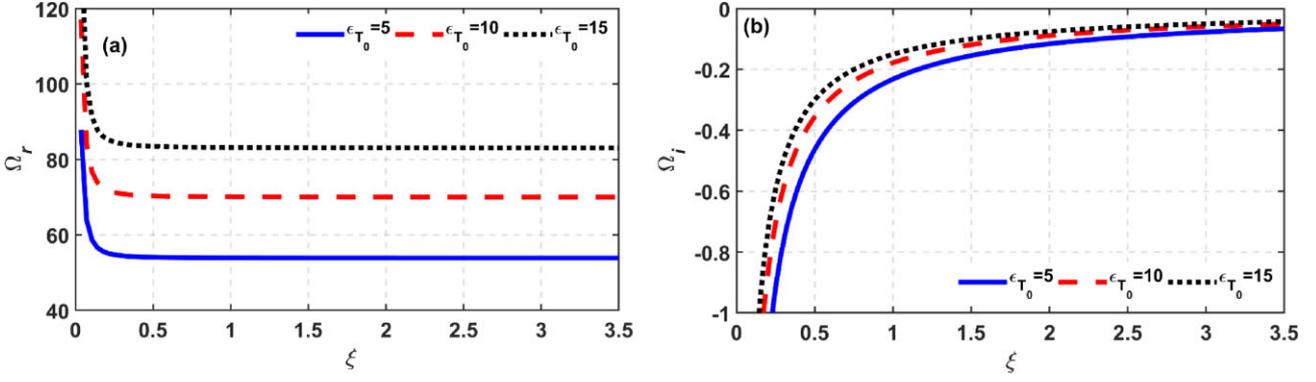

**Figure 7.** Profile of the Jeans-normalized (a) real frequency ($\Omega_r$) and (b) imaginary frequency ($\Omega_i$) with variation in the Jeans normalized radial distance ($\xi$) for different indicated values of the solar core-to-electron temperature ratio ($\epsilon_{T_0}$) associated with the SIP fluctuation dynamics.

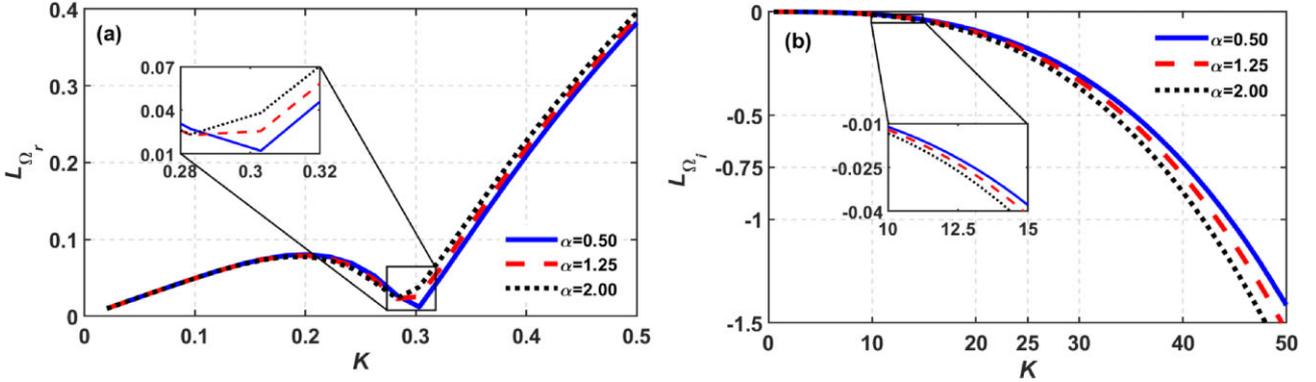

**Figure 8.** Profile of the gradient scale length of (a) real frequency ($L_{\Omega_r}$) and (b) imaginary frequency ($L_{\Omega_i}$) with variation in the Jeans-normalized angular wavenumber ($K$) for different indicated values of the magneto-thermal pressure coupling constant ($\alpha$) associated with the SIP fluctuation dynamics. The microphysical portrayal of the associated transitional behaviours is enlarged in the respective insets.

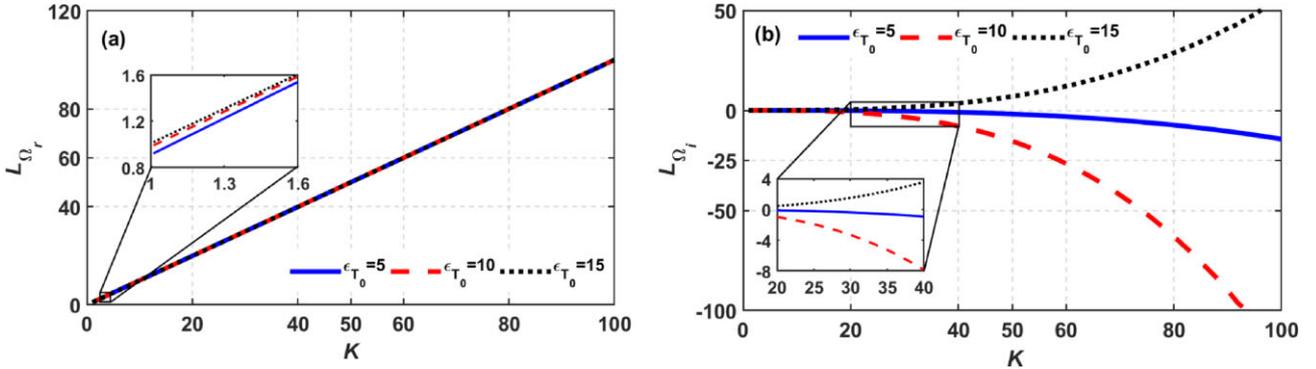

**Figure 9.** Profile of the gradient scale length of (a) real frequency ($L_{\Omega_r}$) and (b) imaginary frequency ($L_{\Omega_i}$) with variation in the Jeans-normalized angular wavenumber ($K$) for different indicated values of the solar core-to-electron temperature ratio ($\epsilon_{T_0}$) associated with the SIP fluctuation dynamics. The microphysical portrayal of the associated transitional behaviours is enlarged in the respective insets.

given as $\Omega_r = f(K)$ for both the short- and long-wave fluctuations (Fig. 4(a)). This kind of non-dispersive characteristics of helioseismic fluctuations arises in the SIP region due to the joint action of non-linear polytropic pressure and strong self-gravitational effects. It is also found that $\Omega_r$ increases less rapidly as $\alpha$ increases, and vice versa. It implies that $\alpha$ works as an accelerating agent for the SIP fluctuations to propagate. Moreover, it also signifies that the average SIP magnetic field, plasma density, and electron temperature affect the SIP fluctuation behaviour. However, the growth rate, $\Omega_i$, is almost uniform; thereby, showing no damping with variation in

$K$ (Fig. 4(b)). It is also noticeable that the magnitude of $\Omega_i$ shows a sharp enhancing tendency as the $\alpha$-value increases and vice versa under the non-local gravito-electrostatic action. It indicates that $\alpha$ plays as a stabilizing factor to this GES-based fluctuations. In other words, one can add herein that, the plasma-$\beta$ parameter plays the role of a destabilizer (decelerating agent) on the SIP regime of the excited collective fluctuations and oscillations (Fig. 4(b)).

In a similar way, Fig. 5 depicts the same as in Fig. 4, but for different indicated values of the solar core-to-electron temperature ratio (e.g. $\epsilon_{T_0} = 5$, 10, 15). It is implicitly noticeable that $\Omega_r$ increases rapidly as the temperature increases, and vice versa (Fig. 5(a)). It indicates that temperature plays as an accelerating agent increasing the group velocity ($d\Omega_r/dK$) of the collective waves on the SIP scale. Moreover, it is also observed that the group velocity of long waves has very slow variation with $\epsilon_{T_0}$ than that of the shorter ones. It could be noticeable that the growth rate of the fluctuations remains almost unchanged with $K$ for constant temperature (Fig. 5(b)). It signifies stable oscillatory propagation of the collective oscillations at the SSB. Furthermore, it is seen that the damping rate of the fluctuations sharply decreases as temperature increases implying the instability propensity of the SIP fluctuations (Fig. 5(b)).

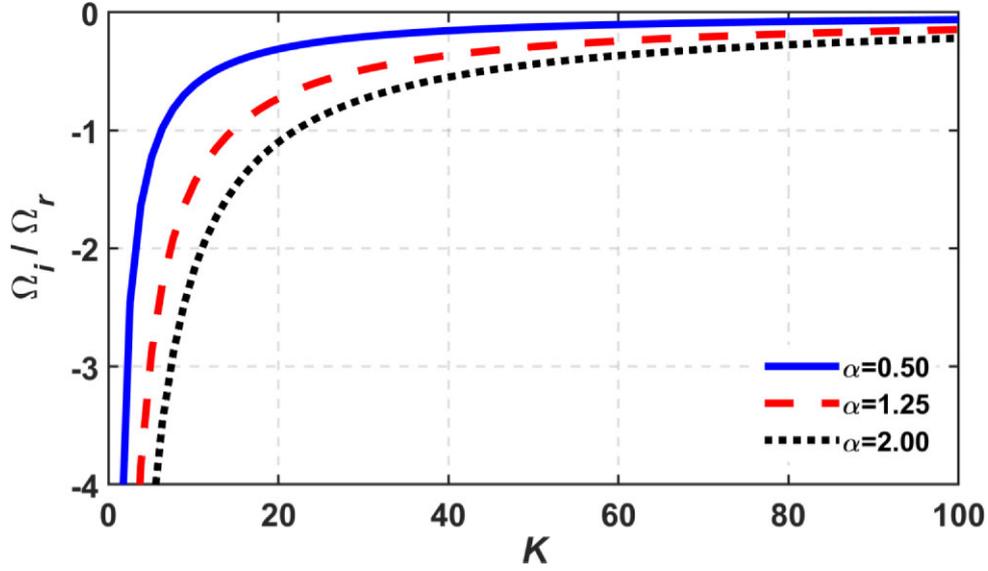

**Figure 10.** Profile of the growth per cycle ($\Omega_i/\Omega_r$) with variation in the Jeans-normalized angular wavenumber ($K$) for different indicated values of the magneto-thermal pressure coupling constant ($\alpha$) associated with the SIP fluctuation dynamics.

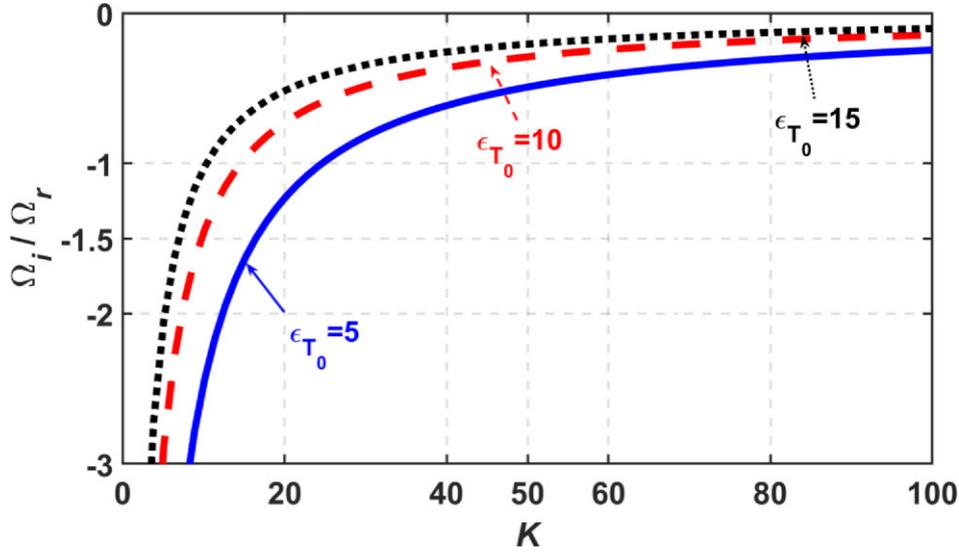

**Figure 11.** Profile of the growth per cycle ($\Omega_i/\Omega_r$) with variation in the Jeans-normalized angular wavenumber ($K$) for different indicated values of the solar core-to-electron temperature ratio ($\epsilon_{T_0}$) associated with the SIP fluctuation dynamics.

In Fig. 6, we exhibit the profile structure of (a) $\Omega_r$ and (b) $\Omega_i$ with variation in $\xi$ for different indicated $\alpha$-values associated with the SIP fluctuation dynamics. We keep the $K$-value fixed at $K = 20$. It is clearly observed that, near the solar core, the SIP collective fluctuations have very high $\Omega_r$-value, thereby indicat-ing its strong propagatory nature. As a consequence, beyond this zone, the fluctuations show stable propagatory nature (Fig. 6(a)). It is also seen that $\Omega_r$ has a slow enhancement nature with increase in the $\alpha$-value. However, for $\xi < 0.5$, the $\Omega_i$-magnitude increases very rapidly with the increasing $\alpha$. It hereby indicates the stabilizing role of $\alpha$ on the collective SIP fluctuation dynamics (Fig. 6(b)).

As in Fig. 7, we depict the same as Fig. 6, but for the different indicated $\epsilon_{T_0}$-values. It shows that $\Omega_r$ increases very rapidly with an enhancement in the $\epsilon_{T_0}$-value. It herewith indicates that the temperature plays the role of accelerating agent on the GES-based fluctuations (Fig. 7(a)). It is also seen that, for $\xi < 1$, the damping rate decreases rapidly with $\epsilon_{T_0}$. It signifies the destabilizing role of the temperature on the SIP fluctuations dynamics (Fig. 7(b)).

In Fig. 8, we show the profile of the gradient scale length of (a) real frequency ($L_{\Omega_r}$) and (b) imaginary frequency ($L_{\Omega_i}$) with variation in $K$ for the different indicated $\alpha$-values associated with the SIP fluctuation dynamics. The respective zoomed-in subplots depict the microphysical details of the collective waves. The growth gradient scale length is given as $L_{\Omega_{r(i)}} = [(1/\Omega_{r(i)})(d\Omega_{r(i)}/dK)]^{-1}$; where $\Omega_{r(i)} =$ real (imaginary) Jeans-normalized frequency and $d\Omega_{r(i)}/dK$ is the growth gradient or group velocity of the collective waves. It is obvious from Fig. 8(a) that for very long-wave fluctuations ($K \sim 0$), $L_{\Omega_r}$ tends to become zero ($L_{\Omega_r} \to 0$), that means the growth gradient or group velocity of the collective waves becomes very high ($d\Omega_r/dK \to \infty$). Up to $K = 0.2$, $L_{\Omega_r}$ increases with the increase of $K$-value. It shows a decreasing propensity with the increasing value of $K$ up to $K = 0.3$. Beyond this scale, for the short-wave fluctuations, it is found interestingly that $L_{\Omega_r}$ increases very sharply with $K$ which is due to the rapid increase of $\Omega_r$-value of the collective waves. Moreover, it is observed that $L_{\Omega_r}$ shows a slight increasing tendency as $\alpha$-value increases (Fig. 8(a)). It is noticeable that for long-wave fluctuations, up to $K = 10$, $L_{\Omega_i}$ has almost no variation with $K$ (Fig. 8(b)). Beyond that region, its magnitude increases gradually lowering the group velocity of the collective waves. Furthermore, it is seen that for short-wave fluctuations, $L_{\Omega_i}$ shows a sharp enhancing tendency as $\alpha$ increases. The $L_{\Omega_i}$-values in our analysis are rescaled by multiplying with $10^{-7}$ for the sake of simplicity of graphical analysis.

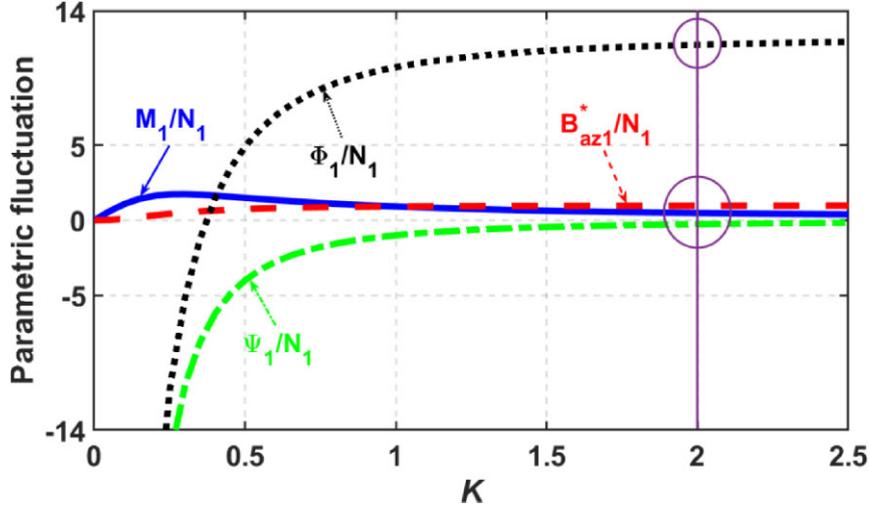

**Figure 12.** Profile of the parametric fluctuation of different perturbed variables with variation in the Jeans-normalized angular wavenumber ($K$) associated with the SIP fluctuation dynamics.

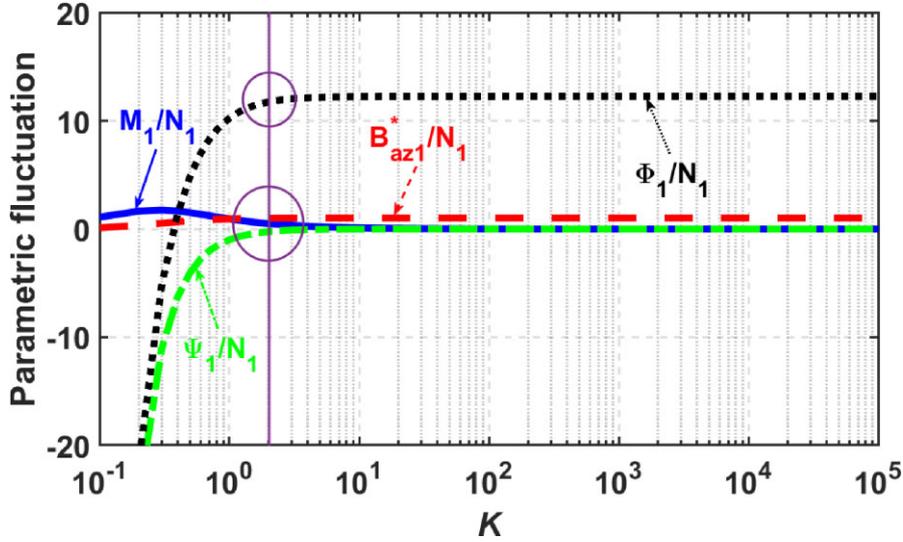

**Figure 13.** Same as Fig. 12, but in the extended $K$-space (high-$K$) on the logarithmic scale.

Fig. 9 portrays same features as shown in Fig. 8, but for different values of $\epsilon_{T_0}$. It is found that $L_{\Omega_i}$ increases sharply with $K$-value. This sharp increasing is due to the very rapid enhancement of $\Omega_r$ with $K$ (Fig. 9(a)). Moreover, $L_{\Omega_r}$ has almost no variation with $\epsilon_{T_0}$. It is noticeable that, up to $K = 20$, the $L_{\Omega_i}$-value is almost zero. Beyond this zone, we can see variation of $L_{\Omega_i}$ with $K$ (Fig. 9(b)). It is also observed that $\epsilon_{T_0}$ shows atypical variation with $\epsilon_{T_0}$. The $L_{\Omega_i}$-values are rescaled by multiplying with $10^{-6}$ for the sake of analytical simplicity without any loss of generality.

In Fig. 10, we depict the profile of the growth per cycle ($\Omega_i/\Omega_r$) with variation in $K$ for different $\alpha$-values associated with the SIP fluctuation dynamics. It is interestingly found that very long-wave fluctuations ($K \sim 0$) have comparatively higher $\Omega_i/\Omega_r$-values, thereby showing dissipative features. As the $K$-value increases, the wave shows more propagating propensity; thereby decreasing the magnitude of $\Omega_i / \Omega_r$ with increasing $K$. Furthermore, for shorter wave, beyond $K > 40$, we see very slow variations and stable propagation behaviour. It is also seen that there exists a critical wavenumber at $K = 10$ which separates the fast and slow variations of the collective wave fluctuations. Moreover, as the $\alpha$-value enhances, $\Omega_i / \Omega_r$ shows an increasing tendency. Hereby, we can anticipate that the dissipative nature of the collective waves enhances with the $\alpha$-value (Fig. 10).

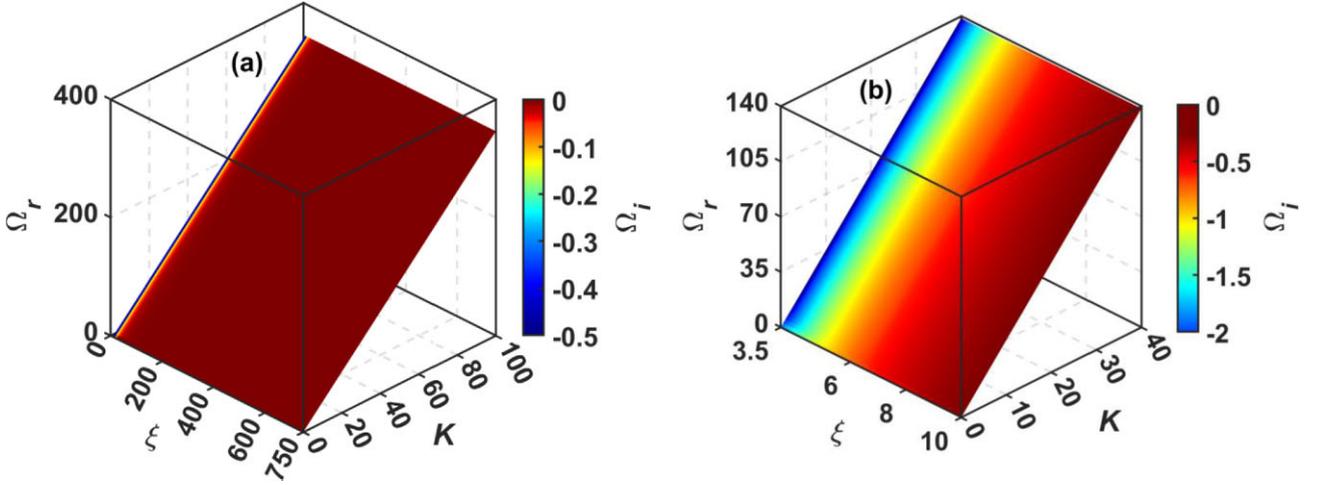

**Figure 14.** Colourspectral profile of the Jeans-normalized real frequency ($\Omega_r$) and imaginary frequency ($\Omega_i$) jointly with variation in the Jeans-normalized radial coordinate ($\xi$) and the Jeans-normalized angular wavenumber ($K$) associated with the SWP fluctuation dynamics (subplot a). The zoomed-in microphysical details are depicted adjacently (subplot b).

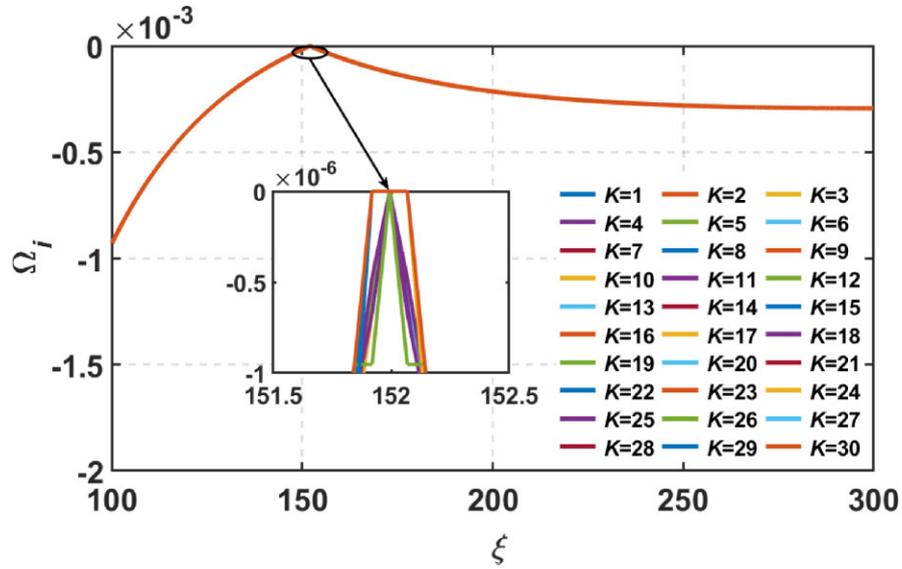

**Figure 15.** Profile of the Jeans-normalized imaginary frequency part ($\Omega_i$) with variation in the Jeans-normalized radial distance ($\xi$) for different indicated values of Jeans-normalized angular wavenumber ($K$) associated with the SWP fluctuation dynamics.

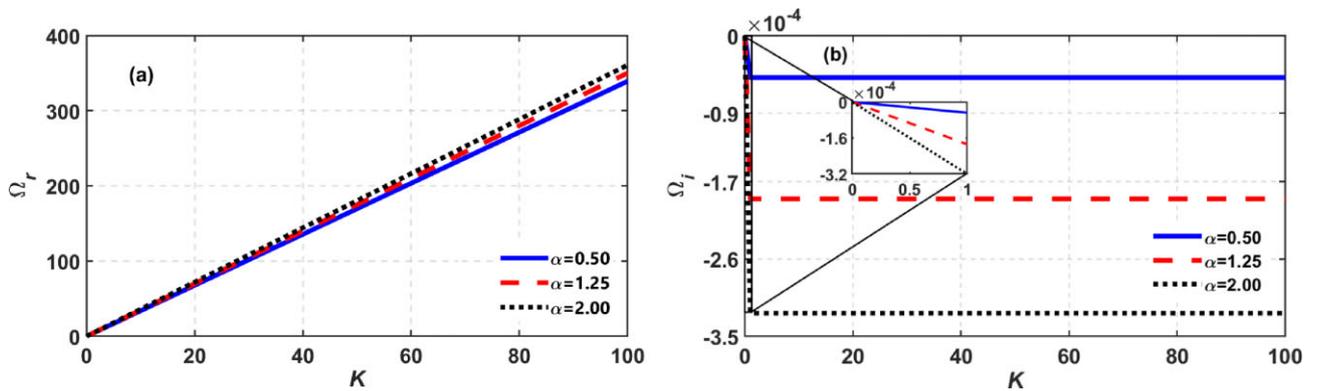

**Figure 16.** Profile of the Jeans-normalized (a) real frequency ($\Omega_r$) and (b) imaginary frequency ($\Omega_i$) with variation in the Jeans-normalized angular wavenumber ($K$) for different indicated values of the magneto-thermal pressure coupling constant ($\alpha$) associated with the SWP fluctuation dynamics.

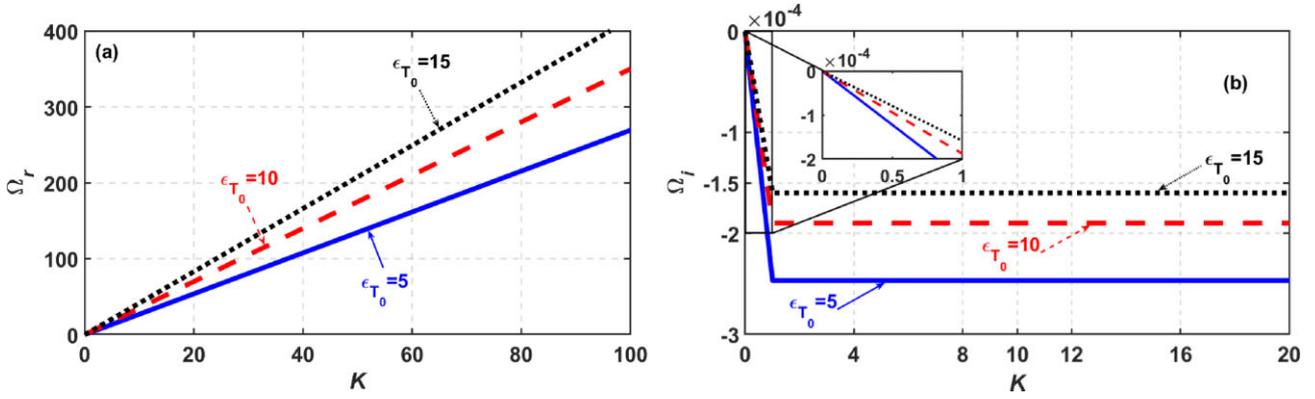

**Figure 17.** Profile of the Jeans-normalized (a) real frequency ($\Omega_r$) and (b) imaginary frequency ($\Omega_i$) with variation in the Jeans-normalized angular wavenumber ($K$) for different indicated values of the solar core-to-electron temperature ratio ($\epsilon_{T_0}$) associated with the SWP fluctuation dynamics.

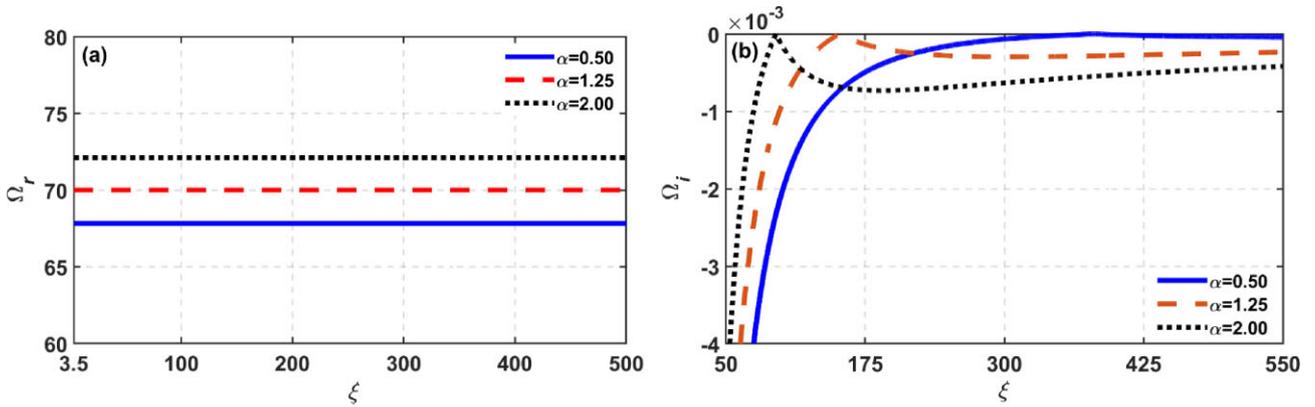

**Figure 18.** Profile of the Jeans-normalized (a) real frequency ($\Omega_r$) and (b) imaginary frequency ($\Omega_i$) with variation in the Jeans normalized radial distance ($\xi$) for different indicated values of the magneto-thermal pressure coupling constant ($\alpha$) associated with the SWP fluctuation dynamics.

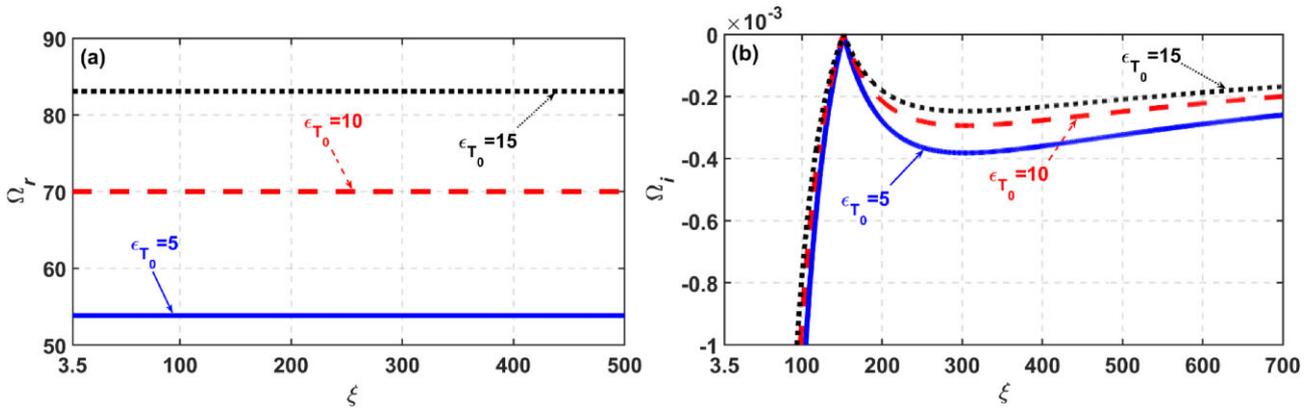

**Figure 19.** Profile of the Jeans-normalized (a) real frequency ($\Omega_r$) and (b) imaginary frequency ($\Omega_i$) with variation in the Jeans normalized radial distance ($\xi$) for different indicated values of the solar core-to-electron temperature ratio ($\epsilon_{T_0}$) associated with the SWP fluctuation dynamics.

Fig. 11 depicts the same as Fig. 10, but for different values of $\epsilon_{T_0}$. It is seen that longer wave fluctuations show comparatively higher $\Omega_i / \Omega_r$-values, thereby indicating dissipative features of the fluctuations. As the $K$-value increases, the wave shows more propagating propensity; thereby decreasing the $\Omega_i / \Omega_r$-value. Furthermore, for shorter wave, beyond $K > 40$, we see very slow variations and stable propagation behaviour. It is also evident that the critical wavenumber, $K = 20$, separates the fast and slow variations of the collective wave fluctuations (Fig. 11). Moreover, as the $\epsilon_{T_0}$-value increases, $\Omega_i / \Omega_r$ shows a decreasing propensity. It can thereby be anticipated that the propagatory nature of the collective waves enhances with the $\epsilon_{T_0}$-value associated with the SIP fluctuation dynamics.

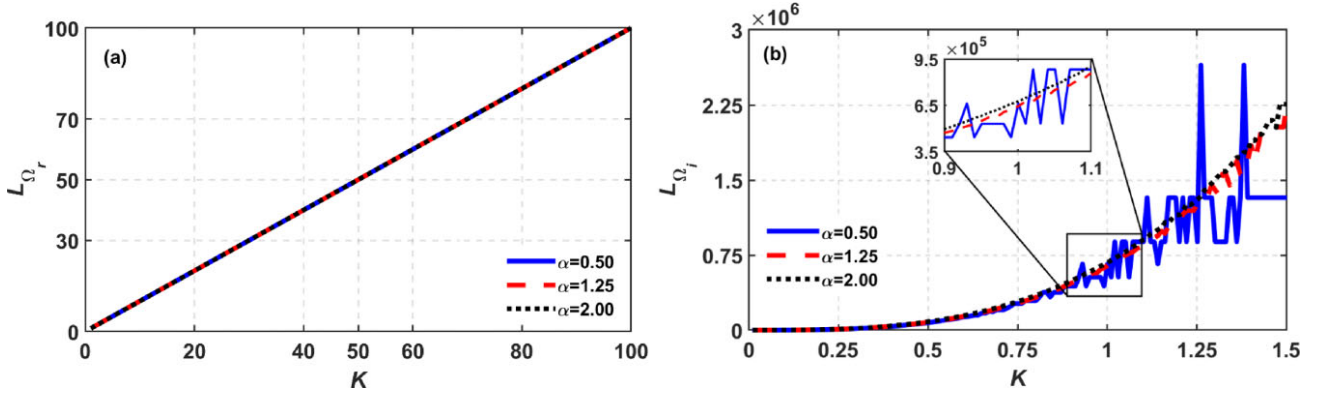

**Figure 20.** Profile of the gradient scale length of (a) real frequency $(L_{\Omega_r})$ and (b) imaginary frequency $(L_{\Omega_i})$ with variation in the Jeans-normalized angular wavenumber $(K)$ for different indicated values of the magneto-thermal pressure coupling constant $(\alpha)$ associated with the SWP fluctuation dynamics. The microphysical portrayal of the associated transitional behaviours is enlarged in the respective insets.

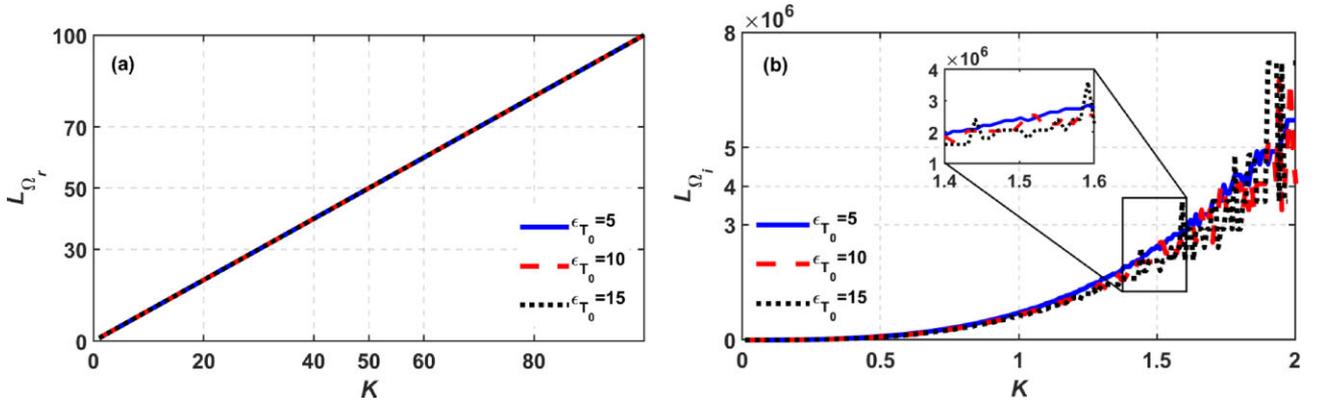

**Figure 21.** Profile of the gradient scale length of (a) Real frequency $(L_{\Omega_r})$ and (b) Imaginary frequency $(L_{\Omega_i})$ with variation in the Jeans-normalized angular wavenumber $(K)$ for different indicated values of the solar core-to-electron temperature ratio $(\epsilon_{T_0})$ associated with the SWP fluctuation dynamics. The microphysical portrayal of the associated transitional behaviours is enlarged in the respective insets.

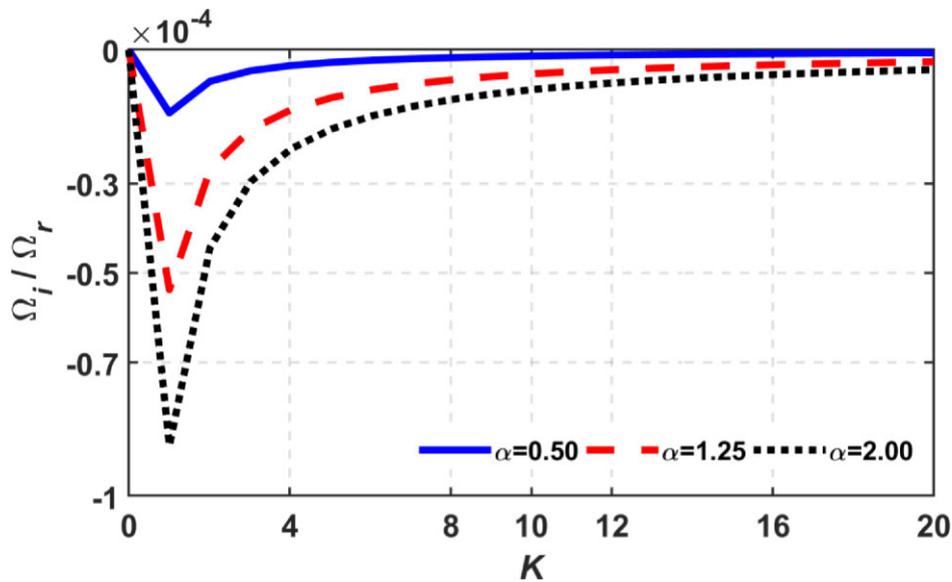

**Figure 22.** Profile of the growth per cycle $(\Omega_i/\Omega_r)$ with variation in the Jeans-normalized angular wavenumber $(K)$ for different indicated values of the magneto-thermal pressure coupling constant $(\alpha)$ associated with the SWP fluctuation dynamics.

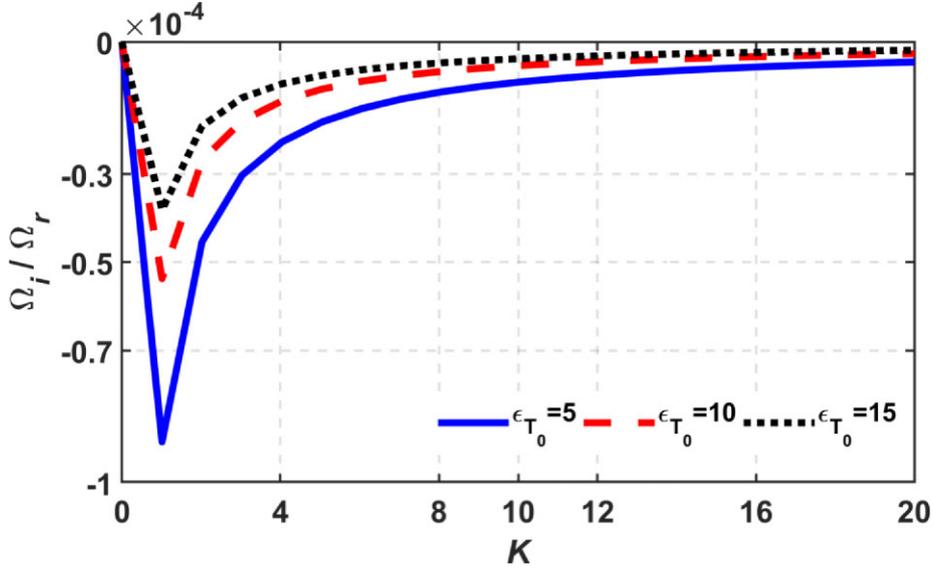

**Figure 23.** Profile of the growth per cycle ($\Omega_i / \Omega_r$) with variation in the Jeans-normalized angular wavenumber ($K$) for different indicated values of the solar core-to-electron temperature ratio ($\epsilon_{T_0}$) associated with the SWP fluctuation dynamics.

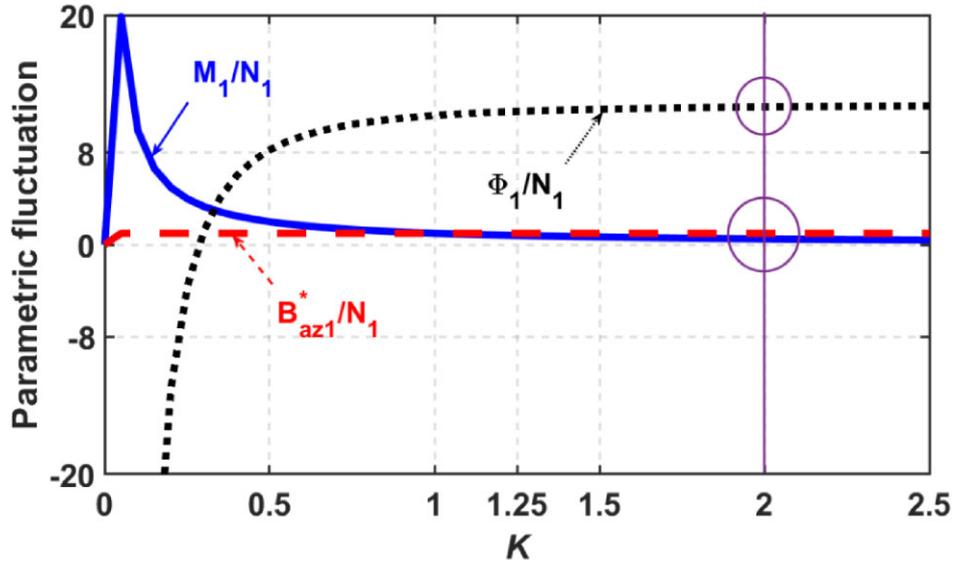

**Figure 24.** Profile of the parametric fluctuation of different perturbed variables with variation in the Jeans-normalized angular wavenumber ($K$) associated with the SWP fluctuation dynamics.

In Fig. 12, we display profile of the parametric fluctuations of different perturbed variables with variation in $K$ associated with the SIP fluctuation dynamics. This graphical presentation is portrayed by performing numerical analysis in equations (20)–(24). The associated parametric fluctuations of the perturbed variables are taken as $M_1/N_1$, $B_{az1}^*/N_1$, $\Phi_1/N_1$, and $\Psi_1/N_1$. It is interestingly seen that $M_1/N_1$ and $B_{az1}^*/N_1$ have values close to zero and hence, show very slow variations with $K$. On the other hand, for longer wave fluctuations ($K < 1$), $\Phi_1/N_1$ and $\Psi_1/N_1$ increase rapidly with the $K$-value. It is also observed that, beyond $K = 2$, all the parametric fluctuations show almost no variation with the $K$-values (Fig. 12).

Fig. 13 depicts the same fluctuation features as shown in Fig. 12, but in the extended $K$-space (high-$K$) on the logarithmic scale. As already analysed above, it is seen that $M_1/N_1$ and $B_{az1}^*/N_1$ are almost zero and have very slow variation with $K$. While, $\Phi_1/N_1$ and $\Psi_1/N_1$ increase rapidly with $K$ for longer waves. In this extended scale, it is again evident that, beyond $K = 2$, all the parametric fluctuations have almost no variation with the increasing $K$-values.

## 3.2 SWP stability outcomes

A detailed numerical analysis is performed based on the Jeans-normalized real frequency part ($\Omega_r$) and imaginary frequency part ($\Omega_i$) expressed, respectively, in equations (32) and (33). It enables us to obtain graphical profiles of the SWP stability dynamics in Figs 14–25. In Fig. 14, we depict the colourspectral 4D profile of the Jeans-normalized real ($\Omega_r$) and imaginary ($\Omega_i$) frequency parts with the variation in the Jeans-normalized radial distance ($\xi$) and Jeans-normalized angular wavenumber ($K$). We show the different zoomed-in subplot corresponding to different spectral scaling. The magneto-thermal pressure coupling constant is kept fixed here as $\alpha = 1.25$. The solar core-to-electron temperature ratio is considered as $\epsilon_{T_0} = 10$. The standard value of the normalization coefficient is taken as $a_0 = 95$. It is clearly observed that in the unbounded SWP scale, the collective waves show purely non-dispersive acoustic-like nature. It holds a linear dispersion relationship given as $\Omega_r = f(K)$. It is pertinent to add here that the fluid turbulence pressure effect becomes so weak in the SWP regime that it cannot employ any wave excitation effect in the wave propagation dynamics. It is noticeable that the $\Omega_r$-value is absolutely constant in the $\xi$-space for any specific $K$-value or wavelength. It indicates a propensity to stable oscillatory propagation of the SWP fluctuations, which is indeed an exact similar fashion observed in the bounded SIP scale. Therefore, we can speculate herewith that the helioseismic collective oscillation modes associated with the non-planar perturbation fluctuation dynamics of the solar plasmas are scale-invariant under the quasi-hydrostatically balanced non-local long-range GES-based force-field action. Moreover, it is found that the perturbation waves are more damped near the SSB than that in the higher radial distance. It is interestingly noticed that $\Omega_i$ remains unchanged with the $K$-value (Fig. 14).

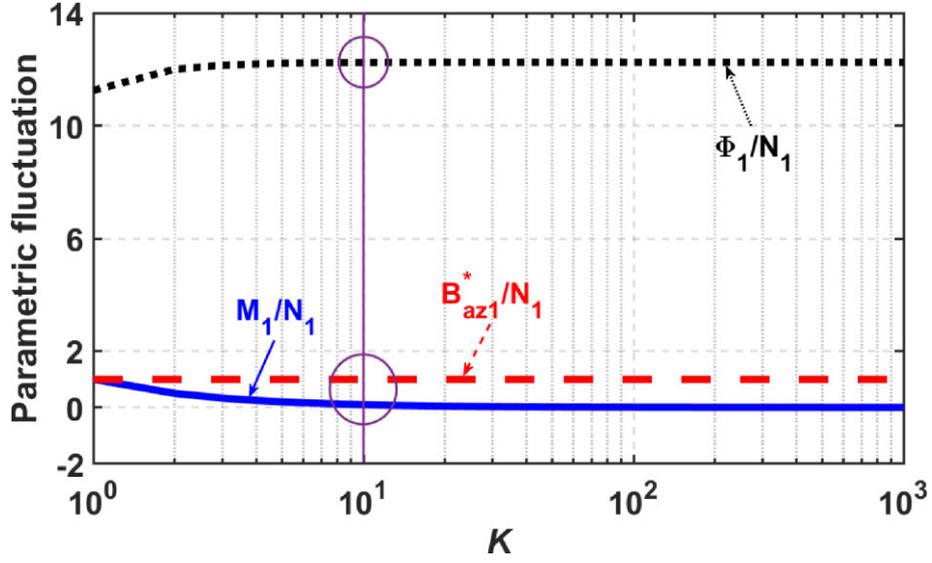

**Figure 25.** Same as Fig. 24, but in the extended $K$-space (high-$K$) on the logarithmic scale.

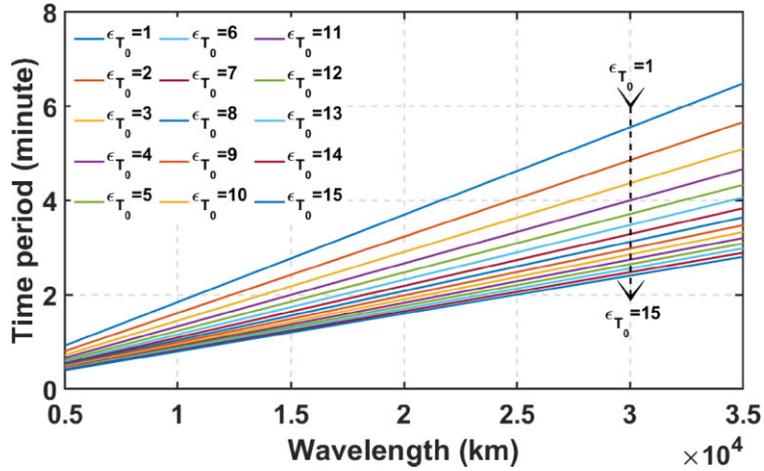

**Figure 26.** Profile of the p-mode time period (in minute) with variation in the horizontal wavelength (in km) for the different indicated values of the solar core-to-electron temperature ratio ($\epsilon_{T_0}$).

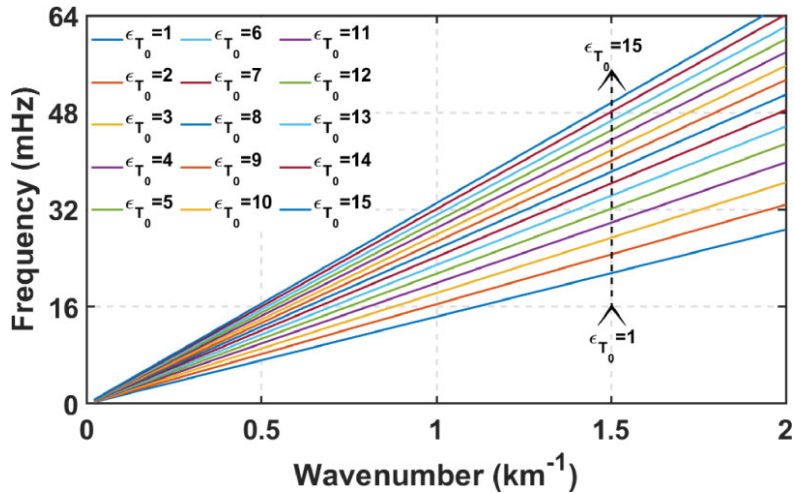

**Figure 27.** Profile of the p-mode frequency (in mHz) with variation in the wavenumber (in km$^{-1}$) for the different indicated values of the solar core-to-electron temperature ratio ($\epsilon_{T_0}$).

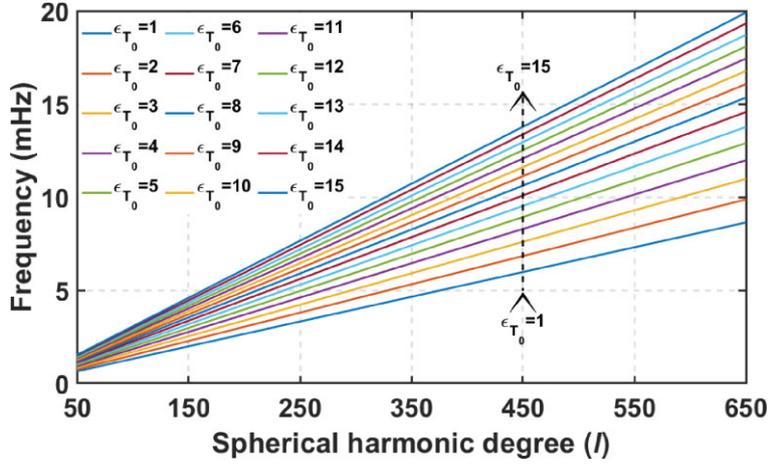

**Figure 28.** Profile of the p-mode wave frequency (in mHz) with variation in the spherical harmonic degree ($l$) for the different indicated values of the solar core-to-electron temperature ratio ($\epsilon_{T_0}$).

In Fig. 15, we display the profile of $\Omega_i$ with variation in $\xi$ for different indicated values of $K$ associated with the SWP fluctuation dynamics. It depicts the damping features of the SWP oscillations. It is evident that $\Omega_i$ is almost constant with $K$. It is found interestingly that, on higher $\xi$-scale ($\xi > 250$), $\Omega_i$ remains uniform. Hence, it ensures a stable oscillatory mode of the collective GES perturbations on the unbounded SWP spatiotemporal regime (Fig. 15).

Fig. 16 depicts the profiles of (a) $\Omega_r$ and (b) $\Omega_i$ with variation in $K$ for different $\alpha$-values for the SWP fluctuation dynamics. We fix the input values here as $a_0 = 95$, $\epsilon_{T_0} = 10$, and $\xi \sim 750$ for a common spectral analysis. It is observed that, similar to the SIP collective waves, the SWP fluctuations exhibit a pure non-dispersive acoustic-like nature providing a linear relationship $\Omega_r = f(K)$. It is also found that in the SWP regime, $\Omega_r$ shows a slight growth trend as $\alpha$ increases, which is similar to that found in the SIP scale (Fig. 16(a)). Moreover, it is evidently noticed that, in the SWP regime, very long-wave oscillations ($K \leq 1$) show higher damping rate than the shorter ones. Beyond $K = 1$, the damping rate remains unchanged with $K$-value. It signifies stable GES-based oscillation modes in the SWP scale. It is observed interestingly that, the damping rate of the SWP fluctuations increases very rapidly with $\alpha$-value and vice versa. It signifies the stabilizing role of $\alpha$ on the SWP regime (Fig. 16(b)).

In Fig. 17, we show the same as Fig. 16, but for different indicated values of $\epsilon_{T_0}$ associated with the SWP fluctuation dynamics. It is found that $\Omega_r$ increases very rapidly with $\epsilon_{T_0}$ (Fig. 17(a)). It indicates that $\epsilon_{T_0}$ plays as a strong accelerating agent for the SWP fluctuations. Moreover, it is seen that the shorter wave fluctuations show more propagatory propensity than the longer ones on the unbounded SWP region. It is explicit that the damping rate of the collective fluctuations decreases rapidly with enhancing $\epsilon_{T_0}$-value. It clearly indicates the strong destabilizing role of $\epsilon_{T_0}$ on the SWP fluctuation dynamics (Fig. 17(b)).

In Fig. 18, we exhibit the profile structure of (a) $\Omega_r$ and (b) $\Omega_i$ with variation in $\xi$ for different indicated $\alpha$-values associated with the SWP fluctuation dynamics. The $K$-value is kept fixed here at $K = 20$. It is seen that $\Omega_r$ remains uniform in $\xi$-scale signifying the stable propagation of the GES-based SWP fluctuations. It is also found that as the $\alpha$-value increases, $\Omega_r$ enhances indicating accelerating role of $\alpha$ (Fig. 18(a)). However, it is observed that for lower $\xi$-value, the collective waves show more damping behaviour than higher radial distance. It is interestingly found that up to $\xi \sim 110$, damping rate decreases with $\alpha$-value; while, for $\xi \geq 235$, the damping rate increases with increasing $\alpha$ and vice versa. Hence, $\alpha$ plays an atypical mixed role on the damping nature of the SWP fluctuation dynamics (Fig. 18(b)).

Fig. 19 depicts same as Fig. 18, but for different indicated values of $\epsilon_{T_0}$. It is seen that an enhancement in $\epsilon_{T_0}$ leads to rapid increment in $\Omega_r$-value. It strengthens the fact that $\epsilon_{T_0}$ plays as strong accelerating agent on the SWP scale (Fig. 19(a)). It is also noticeable that up to $\xi = 150$, the damping rate decreases rapidly with $\xi$; while, for higher $\xi$-value, $\Omega_i$ shows less variation. The damping rate decreases with the $\epsilon_{T_0}$-value; thereby, signifying the destabilizing role of $\epsilon_{T_0}$ on the SWP regime (Fig. 19(b)).

In Fig. 20, we display the profile of (a) $L_{\Omega_r}$ and (b) $L_{\Omega_i}$ with variation in $K$ for different values of $\alpha$ associated with the SWP fluctuation dynamics. The subplot shown here present respective zoomed-in microphysical features of the collective waves in the SWP regime. It is noticed that $L_{\Omega_r}$ shows rapid increasing propensity with increasing $K$. It is because $\Omega_r$ enhances rapidly with $K$. Moreover, it is found that there is no significant variation in $\Omega_r$ with $\alpha$ (Fig. 20(a)). It is interestingly noticeable that, in the SWP scale, for very long-wave fluctuations ($K < 3$), $L_{\Omega_i}$ remains almost constant ($L_{\Omega_i} \sim 0$) with the variation in $K$. Beyond that zone, $L_{\Omega_i}$ shows increasing propensity with $K$. It is found that, for $\alpha = 0.5$, beyond $K = 1$, $L_{\Omega_i}$ shows very random variation with $K$. It is also seen that $L_{\Omega_i}$ has no variation with $\Omega_r$ (Fig. 20(b)).

Fig. 21 portrays same as Fig. 20, but for different indicated values of $\epsilon_{T_0}$. It is found that $L_{\Omega_r}$ has no significant variation with $\epsilon_{T_0}$ (Fig. 21(a)). It is also seen that, up to $K = 0.5$, $L_{\Omega_i}$ is almost zero. Beyond this zone, $L_{\Omega_i}$ increases with the $K$-value. It is interestingly seen that, beyond $K = 1$, $L_{\Omega_i}$ show very atypical random variation with $K$. Moreover, it is seen that $L_{\Omega_i}$ show almost no variation with $\epsilon_{T_0}$-value (Fig. 21(b)).

In Fig. 22, we show the profile of the growth per cycle ($\Omega_i/\Omega_r$) with variation in $K$ for different values of $\alpha$ associated with the SWP fluctuation dynamics. It is observed that, for very long-wave fluctuations, up to $K = 1$, the magnitude of $\Omega_i/\Omega_r$ increases very rapidly with $K$-value. It signifies dissipative nature of the collective GES-based fluctuations on the SWP scale. Beyond this zone, $\Omega_i/\Omega_r$ decreases with $K$-value indicating propagatory propensity of the fluctuations. It is found that, beyond $K = 10$, $\Omega_i/\Omega_r$ has no significant variation with $K$. Moreover, it is also noticeable that, for longer waves, the magnitude of $\Omega_i/\Omega_r$ increases with $\alpha$. Hence, we can speculate that, similar to the SIP scale, in the SWP regime, the dissipative nature of the collective waves enhances with the $\alpha$-value (Fig. 22).

Fig. 23 depicts the same as Fig. 22, but for different values of $\epsilon_{T_0}$. It is seen that, up to $K = 1$, the magnitude of $\Omega_i/\Omega_r$ increases very rapidly with $K$-value indicating dissipative nature of the SWP collective fluctuation dynamics. Beyond that zone, $\Omega_i/\Omega_r$ decreases with $K$; thereby signifying propagatory propensity of the GES-based fluctuations. It is observed that, for $K > 10$, $\Omega_i/\Omega_r$ shows no significant variation with $K$. Moreover, it is noticed that the magnitude of $\Omega_i/\Omega_r$ decreases with $\epsilon_{T_0}$-value. Hence, it can be anticipated herewith that, similar to the SIP scale, on the SWP zone, the propagatory nature of the collective waves and fluctuations enhances with the $\epsilon_{T_0}$-value (Fig. 23).

In Fig. 24, we display profile of the parametric fluctuations of different perturbed variables with variation in $K$ associated with the SWP fluctuation dynamics. This graphical depiction is obtained by performing numerical analysis on perturbed SWP governing equations. The associated parametric fluctuations of the perturbed variables are taken as $M_1/N_1$, $B_{az1}^*/N_1$, and $\Phi_1/N_1$. It is interestingly seen that, $M_1/N_1$ is very close to zero and hence, has very slow variations with $K$ on the SWP zone. $B_{az1}^*/N_1$ remains almost unchanged, with value near to 1, with the $K$-value. It is also observed that $\Phi_1/N_1$ also has very slow variation with $K$. It is interestingly seen that, beyond $K = 1$, there is no significant variation of the parametric fluctuations with $K$ (Fig. 24).

Fig. 25 portrays the same fluctuation features as shown in Fig. 24, but in the extended $K$-space (high-$K$) on the logarithmic scale. It is seen that, for long wave fluctuation, $M_1/N_1$ shows significant variation with $K$. $B_{az1}^*/N_1$ has value close to zero and hence it shows no variation with the $K$-value. Moreover, it is seen that, for $K < 0.5$, $\Phi_1/N_1$ increases rapidly with an enhancement in $K$. It is found interestingly that, beyond $K = 2$, the parametric fluctuations show almost no variation with the increasing $K$-values (Fig. 25).

### 3.3 Atypical illustrated p-mode feature

After performing judicious dispersion analyses on the SIP scale (via equation 26), we investigate the p-mode oscillation characteristics of the helioseismic waves found near the SSB regime. We graphically illustrate the p-mode oscillations characteristics as portrayed in Fig. 26–28. In Fig. 26, we see the dimensional profile of the helioseismic p-mode time period (in minute) with variation in the horizontal wavelength (in km) for the different indicated values of $\epsilon_{T_0}$. It is found interestingly that, the time period increases with the horizontal wavelength, indicating stable p-mode oscillations near the SSB. Moreover, it is speculated that the modal time period decreases with an increase in $\epsilon_{T_0}$ and vice versa. It hereby enables us to infer that $\epsilon_{T_0}$ acts as an accelerating agent towards the propagation of the collective solar plasma waves and oscillations in the regions near the vicinity of the SSB (Fig. 26). It is noteworthy that these investigated diversified p-mode characteristic features, particularly on the wavelength-period correlation, are in good agreement with the standard helioseismic observations made with different astronomic techniques reported previously in the literature (Deubner 1975; Rhodes et al. 1977; Leibacher et al. 1985; Demarque & Guenther 1999; Stix 2002).

As in Fig. 27, we depict the profile of the helioseismic modal p-mode frequency (in mHz) with variation in the wavenumber (in km$^{-1}$) for different indicated values of $\epsilon_{T_0}$. It is seen that the modal frequency shows increasing tendency with an enhancement in the wavenumber signifying stable propagatory propensity of the p-mode oscillations near the SSB region. Furthermore, it is observed that collective frequency very rapidly increases with the $\epsilon_{T_0}$-value and vice versa. It explicitly indicates the accelerating role of the temperature on the helioseismic p-mode fluctuation dynamics (Fig. 27). It is pertinent to add here that these diversified speculations of the wavenumber-frequency features of the p-mode oscillations go fairly in accordance with several solar observations already reported in the literature (Deubner 1975; Rhodes et al. 1977; Demarque & Guenther 1999; Stix 2002).

In Fig. 28, we similarly portray the same as shown in Fig. 27, but with the variation in the spherical harmonic degree ($l$). The degree $l$ of the spherical surface harmonic gives the total number of node circles on the assumed sphere enclosing the solar interior plasma volume. The relationship between the spherical harmonic degree ($l$) and the horizontal modal wavenumber ($k_h$) is cast in a usual form as $k_h R_\odot = [l(l+1)]^{1/2}$ (Christensen-Dalsgaard 2002; Stix 2002; Ambastha 2020). Applying this relation, one can approximate as, $l \simeq k_h R_\odot$ (Christensen-Dalsgaard 2002; Stix 2002; Ambastha 2020). Applying this relation, one can obtain the graphical depiction, as displayed in Fig. 28. It is observed that, the p-mode collective frequency increases with the $l$-value, ensuring stable oscillations near the SSB vicinity. The rest of the modal features are closely the same as already illustrated in Fig. 27. It hereby validates the reliability of our GES-based helioseismic calculation scheme going in fair consistency with the previous *in situ* observational predictions available in the literature (Deubner 1975; Demarque & Guenther 1999; Kosovichev 1999, 2006a; Christensen-Dalsgaard 2002; Stix 2002; Hansen et al. 2004; Gizon & Birch 2005; Ambastha 2020).

In a special case of reduced plane parallel geometry ($\xi \to \infty$, planar), our spherical ($\xi \nrightarrow \infty$, non-planar) analysis shows a fair corroboration with the previously reported planar investigation founded on the same GES-based solar plasma direction (Gohain & Karmakar 2018). It shows a unique feature of the proposed GES-based theory against the diversified sheath formalisms available in the literature (Formisano et al. 1973; Fairfield 1976; Mozer et al. 1978; Gravier et al. 2000; Yamada et al. 2000; Cooling et al. 2001; Oksuz

& Hershkowitz 2005; Alexandrova et al. 2006; Langner et al. 2006; Phan et al. 2007; Richardson et al., 2008, 2022; Siscoe & Odstrcil 2008; Lazarian & Opher 2009; Richardson & Wang 2010; Schoeffler et al. 2011; Wang et al. 2012; Petrinec 2013; Robertson 2013; Lyatsky et al. 2016; Chowdhury et al. 2017; Chasapis et al. 2018; Macek et al. 2018; Rout et al. 2018; Janvier et al. 2019; Shaikh et al. 2020; Ala-Lahti et al. 2021; Sow Mondal et al. 2021). An extensive comparison of the salient parametric features of the modified GES structures of current concern (Dwivedi et al. 2007; Karmakar & Dwivedi 2011; Goutam & Karmakar 2015; Das & Karmakar 2022; Sarma & Karmakar 2022) against various other sheath (non-GES) patterns based on our integrated conceptual fabric for the sake of instant reference is briefly given in Appendix C.

As a quantitative comparison towards authenticity of our work, one can see that the group velocity of our investigated bulk acoustic wave (p mode) is estimated graphically as $v_g = 1.7 \times 10^5$ m s$^{-1}$. The supported normal wave (Alfvén mode) in the solar atmosphere has already been observationally reported with spectroscopic method to propagate with a group velocity of $v_g \sim 1 \times 10^5 - 3 \times 10^5$ m s$^{-1}$ (Hollweg 1978; Berghmans & Clette 1999; Sakurai et al. 2002; Marsh et al. 2003; De Pontieu et al. 2005; McEwan & De Moortel 2006; Antolin & Shibata 2010; Raouafi et al. 2023). This fair velocity matching on the normal bulk acoustic footing from our analysis quantitatively with the results reported in the literature obviously boosts up the reliability of our proposed GES-based theoretic analysis in a simplified way, and so forth.

The different stabilizing and destabilizing agents of physical importance studied in our current investigation are summarily given in Appendix D.

A synoptic highlight of different zone-wise solar plasma characteristic features drawn on the excited diversified collective waves and oscillations investigated in our analyses is concisely provided in a tabular form in Appendix E.

## 4 CONCLUSIONS

In conclusive summary, our theoretic study reports a detailed qualitative, quantitative, and comparative analysis of the helioseismic fluctuation dynamics of solar plasma system founded on the GES model framework on the *Jeansean* spatiotemporal scales. It incorporates the combined action of fluid turbumagnetic pressure, geometrical curvature effect, and magneto-thermal coupling action for the first time. A spherical mode analysis is carried out over the solar governing equations of both the self-gravitationally confined solar plasma (SIP) and its wide-range unbounded atmosphere (SWP). A detailed numerical illustrative analysis characterizes the GES-based solar plasma fluctuation dynamics by introducing a unique pair of generalized quadratic dispersion relations (equations 25 and 31). It is interestingly found here that, the GES-based helioseismic collective fluctuations not only depend on the fluctuation coordinate ($\xi$); but, affected noticeably by the solar core-to-electron temperature ratio ($\epsilon_{T_0}$) and the magneto-thermal pressure coupling constant ($\alpha$). A detailed numerical analysis to reveal different colourspectral profiles (Figs 2–28) is systematically carried out to depict our non-planar constructive analysis based on a standard scale-free *Jeansean* calculation scheme founded on the GES model formalism. We infer that the longer wave (gravitational-like) fluctuations are considerably dispersive, thereby showing more propagatory propensity than the shorter ones (acoustic-like) on the SIP scale (Figs 2–7). In contrast, in the SWP regime, the shorter waves become more propagatory than the longer ones (Figs 14–19). It is also observed that the damping features of the collective fluctuations is much more prominent near the solar core (Fig. 3). Hence, the damping behaviour of the GES-based fluctuations is more explicitly pronounced on the self-gravitational (SIP) scale (Fig. 3) against the acoustic (SWP) scale (Fig. 15). These damping or dissipation processes take place due to the strong influence of the conjoint inhomogeneity effect of the background solar plasmas and solar magnetic field. In our analyses, it is speculated that $\alpha$ plays as a stabilizer (accelerating agent) on the SIP scale (Fig. 6); while, on the SWP scale it shows a mixed role of atypical nature (Fig. 18). Besides, $\epsilon_{T_0}$ plays as a destabilizer (accelerating agent) on both the solar spatiotemporal scales (Figs 7 and 19).

It is pertinent to add here that the quasi-linear inter-coupling behaviour of the gravitational ($K \rightarrow 0$) and the acoustic ($K \rightarrow \infty$) fluctuations theoretically investigated here gives an exact confirmation with the previously reported planar results (Gohain & Karmakar 2018). Thus, our non-planar analysis could hereby provide a validated reliability to carry out further analyses of the GES-based solar plasma fluctuation dynamics in this emerging helioseismic direction of future astroplasmic interest from a new viewpoint of plasma-wall interaction mechanism.

By studying the frequencies and sound speeds of the helioseismic modes, we can infer diverse solar interior characteristic parameters. The major potential applications of the proposed helioseismic analyses in light of the existing solar literatures (Demarque et al. 1994; Demarque & Guenther 1999; Christensen-Dalsgaard 2002; Hansen et al. 2004; Aschwanden 2014; Ambastha 2020; Aerts 2021) could be given as follows:

(1) Theoretical supports to the basic postulates of stellar evolution theory.

(2) Determination of the age of the Sun by helioseismic means as a chronological probe.

(3) Probing the depth of the solar convection zone by means of acoustic speed inversion method as special functions of its vertical depth.

(4) Understanding the solar internal rotation profile.

(5) Detailed probing of the superadiabatic transition layer near the solar surface.

(6) Understanding the solar interior morphodynamics including the deduction of the solar density and temperature profile patterns.

(7) Helioseismic determination of the relative abundance of helium and heavier elements in the entire solar plasma system.

(8) Accurate measurement (*in situ*) of the solar interior magnetic field strength.

(9) Evaluating multipole moments of the solar gravitational potential (quadrupole and higher order moments) by means of helioseismic (oscillation) inversion techniques.

(10) Exploring the G-variability cosmological theory of large-scale structure formation.

(11) Verifying the nature of sterile neutrinos and other potential dark matter candidates modifying the helioseismic mode fluctuation characteristics noticeably.

Our theoretical predictions proposed in this model show fair similarities with many of the astronomical observations reported in the solar literature (Ulrich 1970; Leibacher et al. 1985; Harvey et al. 1996; Hill et al. 1996; Demarque & Guenther 1999; Christensen-Dalsgaard 2002; Hansen et al. 2004; Thompson 2004; Chen & Priest 2006; Kosovichev 2006a, 2006b; Litvinenko & Chae 2017; Duckenfield et al. 2021; Griffiths et al. 2023). A concise glimpse of these conformities is highlighted in a tabular form in Appendix F.

It is well-known that the magnetic reconnection spontaneously takes place in the solar corona and the entire solar plasma system ir-respective of scales (Priest 2014). It develops when distinct magnetic flux systems interact with one another or new flux emerges from beneath the photosphere. This magnetic reconnection mechanism plays a crucial role in the initiation and evolution of myriad solar eruption phenomena, such as the solar flares, coronal mass ejections, and so forth (Priest 2014). We admit herein summarily that all such phenomena driven by magnetic reconnection processes are silent in our proposed GES-based theoretical analysis for the sake of mathematical simplicity. It hereby opens a new scope for future refinement of the GES theory and its practical application in this direction of solar plasma research interest.

It is noteworthy further that the entire solar plasma system shows excessive temperature anisotropy in nature with respect to the ambient magnetic field (Sarfraz et al. 2022; Yoon et al. 2022). This excessive transverse temperature anisotropies yield a significant growth of both electron-cyclotron and electro-mirror instability (Sarfraz et al. 2022) as well as proton-cyclotron and proton-mirror instability (Yoon et al. 2022) in such space plasmas. We can anticipate that, a realistic refinement of the GES model with the inclusion of the velocity momentum-based quasi-linear theory characterized by the velocity and temperature anisotropy of the plasma constitutive species could be a significant futuristic solar plasma problem.

It is widely known that diversified inertial modes, excited in the rotating structures due to the action of Coriolis restoring force, play an important role in redistributing angular momentum and they are sensitive to parameters such as turbulent viscosity, entropy gradient etc. (Gizon et al. 2021; Hanson et al. 2022; Triana et al. 2022; Bhattacharya & Hanasoge 2023). This Coriolis force could be modified by internal magnetic fields, gravity or compressibility, and so forth. Different inertial modes, such as the Rossby waves (arising from Coriolis rotational effects), magneto-Rossby waves (arising from the coupling between Coriolis force and magnetic field), Rossby-gravity waves or Yanai waves (arising from the coupling between Coriolis force and the gravity), have already been reported to exist in the solar plasma system (Gizon et al. 2021; Hanson et al. 2022; Triana et al. 2022). A thorough investigation and exhaustive characterization of diversified solar inertial modes excited in the solar plasma environs holds the promise of providing novel diagnostic capability and physical insight to probe the solar subsurface structure and morphological dynamics through the appropriate process of helioseismic data analyses.

Like the geoseismology, the measuring of travel times and distances of individual acoustic waves in solar plasmas is an important feature yet to be well explored (Duvall et al. 1993; Stix 2002; Ambastha 2020). The study of the time-distance helioseismology is a local helioseismology technique which could enable us to probe the helioseismic studies of the local phenomena, such as subsurface inhomogeneities near sunspots and also help to refine global models of the interior rotational rates and meridional flow velocities in the Sun (Duvall et al. 1993; Zhao 2008). Hence, in the future, considering the time-distance helioseismology, our model could be further refined for better understanding of the solar interior structures and the plasma flow dynamics beneath the sunspots in this solar plasma research directions (Duvall et al. 1993; Zhao 2008; Ambastha 2020).

The high-resolution stellar observations and determination of the stellar fundamental parameters, such as the mass, radius, age, etc., lead to develop different sophisticated diagnostic tools, among which helioseismic inversion method plays a key role (Christensen-Dalsgaard 2002; Buldgen et al. 2022). Helioseismic inversion techniques have been revealed to infer localized internal structure and dynamics of the Sun, including its interior rotational properties, from observed frequencies of solar oscillations and acoustic travel times.

Moreover, inversion of the frequencies and travel times of collective solar helioseismic oscillations could enable us to obtain information about the solar thermodynamics and magnetic properties. It is an important element for understanding the solar constitution and evolution processes (Kosovichev 1999). Using appropriate helioseismic inversion tools, our model could provide a better insight in future to infer solar subsurface dynamics.

Our solar model is purely based on the absolute universality of the Newtonian gravitational coupling constant ($\partial_t G = 0$). It is reported in the literature that only the solar model predictions constructed under the assumption of $\partial_t [\ln(G)] \sim 10^{-12}$ yr$^{-1}$ are consistent with various astronomical observations of the helioseismic modes available in the literature (Guenther et al. 1998). It hereby opens a new scope to validate the consistency and correlation of the proposed GES-based analysis from the Newtonian gravitational viewpoint. Further investigations with the post-Newtonian gravitational formalisms are left here for a future course of studies.

It has recently been confirmed observationally that a wide spectrum of collective oscillatory phenomena, including coronal-loop waves within the solar corona, is dominated by the 5-min oscillation mechanism, which is, indeed, attributable purely to the solar p modes as their potential source (Hindman & Jain 2008; Felipe & Khomenko 2017); thereby, broadening the applicability of our analysis to see the coronal-loop wave dynamics from a plasma-wall interaction view point.

It is widely known that Einstein's general relativity (GR) is extensively applicable to explain diverse astrophysical phenomena at Solar system scale as well as at larger scale, such as clusters, galaxies, etc. (Pani et al. 2011; Cermeño et al. 2019). Later, Eddington-inspired theories become reliable and enduring descendent to Einstein's gravity (Pani et al. 2011). These Eddington-inspired modified gravity theories are well-known to cause strong modifications in the entire solar structure (Casanellas et al. 2012). It immensely affects the evolution and the equilibrium structure of the self-gravitating sun and provides modified core temperature and acoustic mode profiles (Casanellas et al. 2012). In this context, the modern formulation of Eddington-inspired-Born-Infeld (EiBI) theory is extensively explored to present novel modified gravitational description of self-gravitating astrophysical spatiotemporal scales, such as the solar interior structure (Bañados & Ferreira 2010; Casanellas et al. 2012; Yang et al. 2023). Thus, in our solar plasma model, the EiBI theory could be taken fruitfully to explain solar plasma dynamics in the light of the post-Newtonian formalism of modern gravitational scenario. Besides, investigation of the magnetoactive solar wind turbulence sourced in the multifractal perspective sourced in long-range interspecies correlation effects could be an effective study in this context for a better understanding of the entire solar plasma system and helioseismic collective wave-phenomenological features (Gomes et al. 2023).

At this backdrop, we can finally speculate herewith that a futuristic refinement of our model could be implemented in the light of the above-mentioned scenarios of current interest. Also, a better correlation and consistency could be tentatively bridged between the theoretical predictions founded on our proposed GES-model framework and the relevant observational solar data studied by various ongoing and as well as future solar missions (Kasper et al. 2021; Vidotto 2021). It is pertinent to add here that various reliable astronomical data by the *in situ* instruments present in different active spacecrafts, such as NASA-operated Wind, Advanced Composition Explorer (ACE), STEREO, SDO, PSP; and SOHO, Solar Orbiter (SolO) jointly operated by NASA-ESA (Cohen et al. 2021; Kellogg 2022) could be justifiable to corroborate our theoretic investigations paving the way for further futuristic research scope in this emerging helioseismic direction of solar plasma-wall interaction interest. Various payloads onboard Indian Space Research Organisation (ISRO) operating upcoming solar mission, Aditya-L1, are expected to provide reliable multidirectional observations for conceiving major long-lying inaccessible solar plasma problems yet to address well in the future in the current helioseismic fabric (Seetha & Megala 2017; Mayank et al. 2022).

## ACKNOWLEDGEMENTS

The authors gratefully acknowledge the active cooperation availed from the Department of Physics, Tezpur University. The dynamic support of the colleagues of the Astrophysical Plasma and Nonlinear Dynamics Research Laboratory (APNDRL), Department of Physics, and Tezpur University is duly worth mentioning. The financial support received through the SERB Project, Government of India (Grant: EMR/2017/003222), is thankfully recognized. SD would like to wholeheartedly acknowledge the Department of Science and Technology, Government of India, for providing the DST-INSPIRE research fellowship.

## DATA AVAILABILITY

The data underlying this article will be shared on reasonable request to the corresponding author.

## REFERENCES

Aerts C., 2021, Rev. Mod. Phys., 93, 015001
Ala-Lahti M., Dimmock A. P., Pulkkinen T. I., Good S. W., Yordanova E., Turc L., Kilpua E. K. J., 2021, J. Geophys. Res.: Space Phys., 126, e29896
Alexandrova O., Mangeney A., Maksimovic M., Cornilleau-Wehrlin N., Bosqued J.-M., André M., 2006, J. Geophys. Res.: Space Phys., 111, A12208
Aman-Ur-Rehman, Lee J. K., 2018, Phys. Plasmas, 25, 022107
Ambastha A., 2020, Physics of the Invisible Sun. CRC Press, Boca Raton
Antolin P., Shibata K., 2010, 712, 494
Aschwanden M. J., 2014, Encyclopedia of the Solar System, 3rd edn. Elsevier, Boston
Bañados M., Ferreira P. G., 2010, Phys. Rev. Lett., 105, 011101
Berghmans D., Clette F., 1999, Sol. Phys., 186, 207
Bhattacharya J., Hanasoge S. M., 2023, ApJS, 264, 21
Bittencourt J. A., 2004, Fundamentals of Plasma Physics. Springer, New York
Brun A. S., Browning M. K., 2017, Living Rev. Sol. Phys., 14, 4
Buldgen G., Bétrisey J., Roxburgh I. W., Vorontsov S. V., Reese D. R., 2022, Front. Astron. Space Sci., 9, 942373
Casanellas J., Pani P., Lopes I., Cardoso V., 2012, ApJ, 745, 15
Cermeño M., Carro J., Maroto A. L., Pérez-García M. A., 2019, ApJ, 872, 130
Chasapis A. et al. 2018, ApJ, 856, L19
Chen F. F., 1984, Introduction to Plasma Physics and Controlled Fusion. Springer, Boston
Chen P. F., Priest E. R., 2006, Sol. Phys., 238, 313
Chowdhury S., Biswas S., Chakrabarti N., Pal R., 2017, Phys. Plasmas, 24, 062111
Christensen-Dalsgaard J., 2002, Rev. Mod. Phys., 74, 1073
Cohen C. M. S. et al. 2021, A&A, 656, A29
Cooling B. M., Owen C. J., Schwartz S. J., 2001, J. Geophys. Res.: Space Phys., 106, 18763
Das S., Karmakar P. K., 2022, J. Astrophys. Astron., 43, 44
Davood Sadatian S., Gharjeh ghiyaei M., 2021, Adv. Space Res., 68, 3455
Demarque P., Guenther D. B., 1999, Proc. Natl. Acad. Sci., 96, 5356
Demarque P., Krauss L. M., Guenther D. B., Nydam J., 1994, ApJ, 437, 870
De Pontieu B., Erdélyi R., De Moortel I., 2005, ApJ, 624, L61
Deubner F., 1975, A&A, 44, 371
Duckenfield T. J., Kolotkov D. Y., Nakariakov V. M., 2021, A&A, 646, A155
Duvall T. L., Jeffferies S. M., Harvey J. W., Pomerantz M. A., 1993, Nature, 362, 430
Dwivedi C. B., Karmakar P. K., Tripathy S. C., 2007, ApJ, 663, 1340
Fairfield D. H., 1976, Rev. Geophys., 14, 117
Felipe T., Khomenko E., 2017, A&A, 599, L2
Formisano V., Moreno G., Palmiotto F., Hedgecock P. C., 1973, J. Geophys., 78, 3714
García R. A., Ballot J., 2019, Living Rev. Sol. Phys., 16, 4
Gizon L., Birch A. C., 2005, Living Rev. Sol. Phys., 2, 6
Gizon L. et al. 2021, A&A, 652, L6
Gohain M., Karmakar P. K., 2015, Europhys. Lett., 112, 45002
Gohain M., Karmakar P. K., 2018, Results Phys., 8, 592
Gomes L. F., Gomes T. F. P., Rempel E. L., Gama S., 2023, MNRAS, 519, 3623
Goutam H. P., Karmakar P. K., 2015, Astrophys. Space Sci., 357, 127
Goutam H. P., Karmakar P. K., 2016, Europhys. Lett., 115, 29001
Gravier E., Caron X., Bonhomme G., Pierre T., Briançon J. L., 2000, Eur. Phys. J. D, 16, 451
Griffiths M., Gyenge N., Zheng R., Korsós M., Erdélyi R., 2023, Physics, 5, 461
Guenther D. B., Krauss L. M., Demarque P., 1998, ApJ, 498, 871
Gurnett D. A. et al. 2002, Nature, 415, 985
Hale S. J., Chaplin W. J., Davies G. R., Elsworth Y. P., Howe R., 2022, RAS Tech. Instrum., 1, 58
Hansen C. J., Kawaler S. D., Trimble V., 2004, Stellar Interiors. Springer, New York
Hanson C. S., Hanasoge S., Sreenivasan K. R., 2022, Nat. Astron., 6, 708
Harvey J. W. et al., 1996, Science, 272, 1284
Hill F. et al., 1996, Science, 272, 1292
Hindman B. W., Jain R., 2008, ApJ, 677, 769
Hollweg J., 1978, Sol. Phys., 56, 305
Janvier M. et al. 2019, J. Geophys. Res.: Space Phys., 124, 812
Jiulin D., 2006, Astrophys. Space Sci., 305, 247
Jiulin D., 2007, Astrophys. Space Sci., 312, 47
Karmakar P. K., Dwivedi C. B., 2011, Int. J. Astron. Astrophys., 01, 210
Karmakar P. K., Goutam H. P., Lal M., Dwivedi C. B., 2016, MNRAS, 460, 2919
Kasper J. C. et al. 2021, Phys. Rev. Lett., 127, 255101
Kellogg P. J., 2003, ApJ, 925, 106
Kosovichev A. G., 1999, J. Comput. Appl. Math., 109, 1
Kosovichev A. G., 2006a, Adv. Space Res., 37, 1455
Kosovichev A. G., 2006b, Adv. Space Res., 38, 876
Langner U. W., Potgieter M. S., Fichtner H., Borrmann T., 2006, ApJ, 640, 1119
Lazarian A., Opher M., 2009, ApJ, 703, 8
Leibacher J. W., Noyes R. W., Toomre J., Ulrich R. K., 1985, Sci. Am., 253, 48
Leighton R. B., Noyes R. W., Simon G. W., 1962, ApJ, 135, 474
Litvinenko Y. E., Chae J., 2017, A&A, 599, A15
Livadiotis G., 2015, Entropy, 17, 2062
Livadiotis G., McComas D. J., 2009, J. Geophys. Res.: Space Phys., 114, A11105
Livadiotis G., McComas D. J., 2013, Space Sci. Rev., 175, 183
Lyatsky W., Pollock C., Goldstein M. L., Lyatskaya S., Avanov L., 2016, J. Geophys. Res.: Space Phys., 121, 7713
Macek W. M., Wawrzaszek A., Kucharuk B., 2018, Nonlinear Process Geophys., 25, 39
McEwan M. P., De Moortel I., 2006, A&A, 448, 763
Mani P., Hanson C. S., Hanasoge S., 2022, ApJ, 926, 127
Marsh M. S., Walsh R. W., De Moortel I., Ireland J., 2003, A&A, 404, L37
Mayank P., Vaidya B., Chakrabarty D., 2022, ApJS, 262, 23
Mozer F. S., Torbert R. B., Fahleson U. V., Fälthammar C.-G., Gonfalone A., Pedersen A., 1978, Advances in Magnetospheric Physics with GEOS-1 and ISEE. Springer, Dordrecht, p. 471
Narita Y., 2012, Plasma Turbulence in the Solar System. Springer, Berlin
Oksuz L., Hershkowitz N., 2005, Plasma Sources Sci. Technol., 14, 201

Pani P., Cardoso V., Delsate T., 2011, Phys. Rev. Lett., 107, 031101

Parker E. N., 1958, ApJ, 128, 664

Petrinec S. M., 2013, Terr. Atmos. Ocean Sci., 24, 265

Phan T. D. et al. 2007, Geophys. Res. Lett., 34, L14104

Priest E., 2014, Magnetohydrodynamics of the Sun. Cambridge Univ. Press, Cambridge

Raouafi N. E. et al. 2023, ApJ, 945, 28

Rhodes E. J. Jr., Ulrich. R. K., Simon G. W., 1977, ApJ, 218, 901

Richardson J. D., Wang C., 2010, ApJ, 711, L44

Richardson J. D., Kasper J. C., Wang C., Belcher J. W., Lazarus A. J., 2008, Nature, 454, 63

Richardson J. D., Burlaga L. F., Elliott H., Kurth W. S., Liu Y. D., von Steiger R., 2022, Space Sci. Rev., 218, 35

Robertson S., 2013, Plasma Phys. Control Fusion, 55, 093001

Rout D. et al. 2018, J. Geophys. Res.: Space Phys., 123, 4298

Rubab N., Murtaza G., 2006, Phys. Scr., 74, 145

Sakurai T., Ichimoto K., Raju K. P., Singh J., 2002, Sol. Phys., 209, 265

Sarfraz M., López R. A., Ahmed S., Yoon P. H., 2022, MNRAS, 509, 3764

Sarma P., Karmakar P. K., 2022, J. Astrophys. Astron., 43, 46

Schoeffler K. M., Drake J. F., Swisdak M., 2011, ApJ, 743, 70

Seetha S., Megala S., 2017, Curr. Sci., 113, 610

Shaikh Z. I., Raghav A. N., Vichare G., Bhaskar A., Mishra W., 2020, MNRAS, 494, 2498

Siscoe G., Odstrcil D., 2008, J. Geophys. Res.: Space Phys., 113, A00B07

Sow Mondal S., Sarkar A., Vaidya B., Mignone A., 2021, ApJ, 923, 80

Stix M., 2002, The Sun. Springer, Berlin

Thompson M. J., 2004, Astron. Geophys., 45, 4.21

Triana S. A., Guerrero G., Barik A., Rekier J., 2022, ApJ, 934, L4

Ulrich R. K., 1970, ApJ, 162, 993

Vazquez-Semadeni E., Canto J., Lizano S., 1998, ApJ, 492, 596

Vidotto A. A., 2021, Living Rev. Sol. Phys., 18, 3

Wang C. P., Gkioulidou M., Lyons L. R., Angelopoulos V., 2012, J. Geophys. Res.: Space Phys., 117, A08215

Yamada M., Ji H., Hsu S., Carter T., Kulsrud R., Trintchouk F., 2000, Phys. Plasmas, 7, 1781

Yang Q., Tan L., Chen H., Liu S., 2023, New Astron., 99, 101947

Yoon P. H., Sarfraz M., Ali Z., Salem C. S., Seough J., 2022, MNRAS, 509, 4736

Zhao J., 2008, Adv. Space Res., 41, 838

# APPENDIX A: ADOPTED NOTATIONAL SYMBOLISM

Various standard relevant notations and symbols associated with the adopted astrophysical normalization scheme for solar plasma description are highlighted as follows:

| Physical parameter | Symbol | Magnitude |
|---|---|---|
| Electron mass | $m_e$ | $9.31 \times 10^{-31}$ kg |
| Ion mass | $m_i$ | $1.67 \times 10^{-27}$ kg |
| Electron charge | $q_e = -e$ | $-1.6 \times 10^{-19}$ C |
| Ion charge | $q_i = +e$ | $+1.6 \times 10^{-19}$ C |
| Electron temperature | $T_e$ | $10^2$ eV |
| Ion temperature | $T_i$ | $10$ eV |
| Solar core temperature | $T_0$ | $10^3$ eV |
| Jeans frequency | $\omega_J = c_s/\lambda_J$ | $10^{-3}$ s |
| External gravity-rescaling constant | $a_0 = GM_\odot/c_s^2\lambda_J$ | $95$ |
| Universal gravitational constant | $G$ | $6.67 \times 10^{-11}$ m$^3$ kg$^{-1}$ s$^{-2}$ |
| Mean solar mass | $M_\odot$ | $2 \times 10^{30}$ kg |
| Magneto-thermal pressure coupling constant or parameter | $\alpha = B_0^2/\mu_0 n_0 T_e = 2/\beta$ | Subcritical ($\sim 10^{-16}$) |
| Plasma-$\beta$ parameter | $\beta = 2\mu_0 n_0 T_e/B_0^2 = 2/\alpha$ | Hypercritical ($\sim 10^{16}$) |

# APPENDIX B: ADOPTED ASTROPHYSICAL NORMALIZATION SCHEME

The adopted standard astrophysical normalization scheme together with typical values of the respective normalizing parameters in all the customary *Jeansean* notations are cast as follows:

| Normalized parameter | Normalizing parameter | Magnitude |
|---|---|---|
| Radial distance ($\xi = r/\lambda_J$) | Jeans length ($\lambda_J$) | $3.1 \times 10^8$ m |
| Time ($\tau = t/\omega_J^{-1}$) | Jeans time ($\omega_J^{-1}$) | $10^3$ s |
| Population density ($N_{e(i)} = n_{e(i)}/n_0$) | Mean SIP density ($n_0$) | $10^{30}$ m$^{-3}$ |
| Mach number ($M_{e(i)} = v_{e(i)}/c_s$) | Sound phase speed ($c_s$) | $3.1 \times 10^5$ m s$^{-1}$ |
| Gravitational potential ($\Psi = \psi/c_s^2$) | Sound phase speed squared ($c_s^2$) | $9.5 \times 10^{10}$ m$^2$ s$^{-2}$ |
| Electrostatic potential ($\Phi = e\phi/T_e$) | Thermal potential ($T_e/e$) | $10^2$ J C$^{-1}$ |
| Magnetic field ($B_{az}^* = B_{az}/B_0$) | Average SIP magnetic field ($B_0$) | $10^{-4}$ T |

## APPENDIX C: COMPARISON OF GES WITH OTHER SHEATHS

A broad comparison of the active saliant parametric features of the GES structures against the existing various other sheath (non-GES) patterns based on our integrated conceptional understanding for the sake of instant visualization of its readers is presented as follows:

| S No | Item | Laboratory sheath | Magnetosheath | Heliosheath | ICME sheath | GES |
|------|------|-------------------|---------------|-------------|-------------|-----|
| 1 | Degree of neutrality | Non-neutral | Quasi-neutral | Quasi-neutral | Quasi-neutral | Quasi-neutral |
| 2 | Bohm criterion | Yes | NA | NA | NA | Yes, equivalently |
| 3 | Value of Mach number | $M > 1$ | $M \geq \sqrt{2}$ | $M \geq \sqrt{2}$ | $M \geq \sqrt{2}$ | $M \geq \sqrt{2}$ |
| 4 | Existential scale | Debye length | Solar radial length | Solar radial length | Solar radial length | Jeans scale length |
| 5 | Thermal stability | More | Less | Very less | Less | Under investigation |
| 6 | Eigenmodes | Acoustic mode (heavy) | Alfvén | Alfvén | Alfvén | GES-modes |
| 7 | Formalism | Kinetic and fluid | Kinetic and fluid | Kinetic and fluid | Kinetic and fluid | Fluid only |
| 8 | Turbulence effect | NA | Yes | Yes | Yes | Yes |
| 9 | Magnetic field | Absent | Present | Present | Present | Present |
| 10 | MHD applicability | No | Yes | Yes | Yes | No |
| 11 | Electron energy | $2.00 - 3.00$ eV | $1.00 \times 10^2 - 2.00 \times 10^2$ eV | $1.00 \times 10$ eV | $4.00 \times 10^2$ keV | $1.00 \times 10^2$ eV |
| 12 | Temperature | $1.00 \times 10^4$ K | $1.00 \times 10^6$ K | $1.00 \times 10^5$ K | $1.00 \times 10^9$ K | $1.00 \times 10^6$ K |
| 13 | Gravitothermal coupling | Unexplored | Unexplored | Unexplored | Unexplored | Explored |
| 14 | Effect of collision | Pressure dependent | Weakly collisional | Collisionless | Collisionless | Collisionless |
| 15 | Thermostatistical distribution law | Maxwellian | Non-Maxwellian | Non-Maxwellian | Non-Maxwellian | Both Maxwellian and non-Maxwellian |
| 16 | Extensivity/Non-extensivity features | Extensive | Non-extensive | Non-extensive | Non-extensive | Non-extensive |
| 17 | q-entropy | $q \rightarrow 1$ | $q < 1$ (superextensive) $q > 1$ (subextensive) | $q < 1$ | $q < 1$ | $q < 1$ |
|    |          |        |               | $q > 1$ | $q > 1$ | $q > 1$ |
| 18 | Transonic width | Subsonic | Supersonic | Supersonic | Supersonic | Subsonic |
| 19 | Plasma density | $1.00 \times 10^7$ cm$^{-3}$ | $1.00 \times 10$ cm$^{-3}$ | $1.00 \times 10^{-3}$ cm$^{-3}$ | $1.00 \times 10^2$ cm$^{-3}$ | $1.00 \times 10^{30}$ cm$^{-3}$ |
| 20 | Effect of gravity | Microgravity (Newtonian) | External gravity (Newtonian) | External gravity (Newtonian) | External gravity (Newtonian) | Self-gravity (non-Newtonian) |
| 21 | Effect of dust presence | Yes | Yes | Yes | Yes | Yet to be reported |
| 22 | Effect of background wind | NA | Yes (Solar wind) | Yes (Solar wind) | Yes (Solar wind) | NA |
| 23 | Propagatory feature | No | No | No | Yes | No |
| 24 | Expansive feature | Yes | Yes | Yes | No | Yet to be explored |
| 25 | Electric field strength | $1.00 \times 10^{-2}$ V m$^{-1}$ | $1.00 \times 10^{-3}$ V m$^{-1}$ | $1.00 \times 10^{-6} - 1.00 \times 10^{-5}$ V m$^{-1}$ | $1.00 \times 10^{-3}$ V m$^{-1}$ | $-1.00 \times 10^{-7}$ V m$^{-1}$ |
| 26 | Magnetic field strength | $4.00 \times 10^{13}$ nT | $1.00 \times 10$ nT | $1.00 \times 10^{-1}$ nT | $5.00 \times 10$ nT | $1.00 \times 10^5$ nT |
| 27 | Effect of magnetic field | Insignificant in longitudinal direction | Significant in transverse direction | Unexplored | Sheath thickness is affected | SSB expansion |
| 28 | Magneto-thermal pressure coupling | Unexplored | Unexplored | Unexplored | Unexplored | Stabilizes plasma fluctuation |
| 29 | Current density | $1.00 \times 10^6$ A m$^{-2}$ | $1.50 \times 10^{-6}$ A m$^{-2}$ | $1.15 \times 10^2$ A m$^{-2}$ | $7.00 \times 10^{-9}$ A m$^{-2}$ | $6.34 \times 10^{17}$ A m$^{-2}$ |
| 30 | Divergence-free current | Unexplored | Unexplored | Unexplored | Unexplored | Find everywhere except SSB |
| 31 | Bulk velocity at sheath entrance | $\geq c_s$ | $4.50 \times 10^2$ km s$^{-1}$ | $1.50 \times 10^2$ km s$^{-1}$ | $4.50 \times 10^2 - 1.00 \times 10^3$ km s$^{-1}$ | $3.00 \times 10^{-5}$ km s$^{-1}$ |
| 32 | Skin depth | 5.30 m | $5.30 \times 10^4$ m | $7.20 \times 10^6$ m | $7.20 \times 10^4$ m | $5.70 \times 10^{-11}$ m |

## APPENDIX D: DIFFERENT STABILIZER AND DESTABILIZER PARAMETERS

A concise summary of the various stabilizing and destabilizing agents investigated in our study can be presented in a tabular form as follows:

| S No | Parameter | SIP | SWP |
|------|-----------|-----|-----|
| 1 | Magneto-thermal pressure coupling constant | Stabilizer (accelerating agent) | Mixed |
| 2 | Plasma-$\beta$ parameter | Destabilizer (decelerating agent) | Mixed (accelerating agent) |
| 3 | Magnetic field | Stabilizer (accelerating agent) | Mixed (accelerating agent) |
| 4 | Temperature | Destabilizer (accelerating agent) | Destabilizer (accelerating agent) |
| 5 | Equilibrium number density | Destabilizer (decelerating agent) | Mixed (decelerating agent) |

## APPENDIX E: SPATIAL WAVE FEATURES

An itemized synopsis of the different characteristic features of the diverse collective waves and oscillations found in different solar plasma zones can be summarily portrayed as follows:

| Wave type | Item | Solar plasma region | | |
|-----------|------|---------------------|---|---|
| | | SIP ($\xi < 3.5$) | SSB ($\xi = 3.5$) | SWP ($\xi > 3.5$) |
| Long-wavelength (Gravitational) | Propagating nature | Very propagatory | Very less propagatory | Non-propagatory |
| | Dispersive nature | Dispersive | Non-dispersive | Non-dispersive |
| | Damping nature | Less damped | Less damped | Very less damped |
| Short-wavelength (Acoustic) | Propagating nature | Non-propagatory | More propagatory | Very propagatory |
| | Dispersive nature | Non-dispersive | Non-dispersive | Non-dispersive |
| | Damping nature | More damped | Less damped | Less damped |

## APPENDIX F: OUR RESULTS IN SUPPORT OF ASTRONOMIC OBSERVATIONS

A comprehensive glimpse of our investigated results in support of the previously reported diverse astronomic observations in the helioseismic backdrop may be highlighted as follows:

| S No | Item | SSM prediction | Our prediction | Remark |
|------|------|----------------|----------------|--------|
| 1 | Group velocity of bulk acoustic mode (at SSB) | $100 - 300$ km s$^{-1}$ | $170$ km s$^{-1}$ | Our results give the mean modal velocity which fairly matches with diverse astronomic observations |
| 2 | Frequency ($\nu$)-wavenumber ($k$) relation | $\nu \approx 1.5 - 32$ mHz with $0 \leq k \leq 1$ | $\nu \approx 1.5 - 32$ mHz with $0 \leq k \leq 1$ | Our results go fairly in accordance with diverse astronomic observations |
| 3 | Time period ($T$)-wavelength ($\lambda$) relation | $T \approx 0 - 5$ min with $0 \leq \lambda \leq 30000$ km | $T \approx 0 - 6$ min with $0 \leq \lambda \leq 30000$ km | Our analyses go fairly in accordance with observations |
| 4 | Frequency ($\nu$)-spherical harmonic degree ($l$) relation | $\nu \approx 1.5 - 5$ mHz with $0 \leq l \leq 150$ | $\nu \approx 1.5 - 5$ mHz with $0 \leq l \leq 150$ | Our results fairly match with astronomic observations |
| 5 | Group velocity of bulk acoustic mode with temperature | Increases with temperature | Increases with temperature but less rapidly | Our results show similarity with diverse observations |
| 6 | Propagation with temperature | Become more propagatory as temperature increases | Become more propagatory as temperature increases on both scales | Our results are reliable on the grounds of diverse astronomic observational data |
| 7 | Modal damping with temperature | Sound waves show less damping propensity with increasing temperature | Waves show less damping propensity with increasing temperature on both scale | Our results support previous observational predictions |
| 8 | Group velocity of bulk acoustic mode with magnetic field | Direct proportional to magnetic field | Increases with magnetic field but very less rapidly | In accordance with various observational data |
| 9 | Group velocity of bulk acoustic mode with plasma number density | Inversely proportional to square root of plasma number density | Decreases with plasma number but very less rapidly | Qualitative similarities with previous observational predictions |
| 10 | Modal propagation at SSB | Shorter waves are more propagatory than longer waves near SSB | Qualitatively same | Our investigations are supported by previous astronomic observations |
| 11 | Non-dispersion at SSB | Waves show non-dispersive tendency near SSB | Waves become non-dispersive at SSB | It is as per diverse astronomic observations |
| 12 | Modal propagation near core | Longer waves are more propagatory than shorter waves near core | Qualitatively same | Our results show similarities with observations |
| 13 | Intense dispersion near core | Waves become dispersive near core | Waves show more dispersive behaviour near the core | Our results match with various observational data |
| 14 | Dispersion-wavelength variation near the core | Longer waves are more dispersive than shorter ones core-wards | Qualitatively same | Our analyses are reliable with observational support |
| 15 | Propagation in SWP | Shorter waves are more propagatory than the longer waves | Shorter waves are more propagatory than longer waves | Our results match fairly with observations |
| 16 | Dispersion-wavelength variation at SWP | Dispersive for long-wavelength and non-dispersive for short-wavelength | Qualitatively same | Our results show fair corroboration with observations |
| 17 | Modal damping with wavelength in SWP | Shorter acoustic waves are very strongly damped | Shorter waves show more damping propensity than the longer ones | Our analyses are in fair agreement with previously reported solar observations |
| 18 | Modal damping with magnetic field | Magnetic field leads to rapid damping of magnetoacoustic waves | Increasing magnetic field stabilizes SIP fluctuations | Our results match with astronomic observations |
| 19 | Modal damping with plasma-$\beta$ parameter | Increase in plasma-$\beta$ lessens damping rate (destabilizing agent) | Increasing plasma-$\beta$ destabilizes SIP fluctuations | Our predictions are astronomically fairly supported |
| 20 | Modal damping with plasma population density | Damping rate decreases with plasma number density | Solar plasma destabilizes with increasing plasma number density | Our analyses fairly corroborate with previously reported astronomic observations |

This paper has been typeset from a Microsoft Word file prepared by the author.